\documentclass[review, a4paper,fleqn, 5p]{cas-dc}

\usepackage{mdframed}
\usepackage{epstopdf}
\usepackage[caption=false]{subfig}
\usepackage{graphicx}
\usepackage{comment}
\usepackage{xcolor}
\usepackage{multirow, makecell}			
\usepackage{booktabs}					
\usepackage{tabularx}					
\usepackage{threeparttable}				
\usepackage{siunitx}
\usepackage{enumitem, xspace}
\usepackage{mdframed}
\usepackage{array}
\usepackage{flushend}
\newcolumntype{P}[1]{>{\raggedright\arraybackslash}p{#1}}

\newcommand{\faktisk}[0]{\textsc{Faktisk}\xspace}
\newcommand{\faktiskbar}[0]{\textsc{Faktisk Verifiserbar}\xspace}

\newcommand{\anonycountry}[0]{\textsc{Norway}\xspace}

\definecolor{framecolor}{rgb}{0.5, 0.0, 0.13} 

\mdfdefinestyle{mystyle}{ 
    linecolor=framecolor,
    linewidth=1pt,
    backgroundcolor=white,
    roundcorner=10pt,
}
\newmdenv[style=mystyle]{modernframe} 

\usepackage[numbers]{natbib}

\def\tsc#1{\csdef{#1}{\textsc{\lowercase{#1}}\xspace}}
\tsc{WGM}
\tsc{QE}

\begin{document}
\let\WriteBookmarks\relax
\def\floatpagepagefraction{1}
\def\textpagefraction{.001}

\shorttitle{}    

\shortauthors{Khan \emph{et al.}}  

\title [mode = title]{Online Multimedia Verification with Computational Tools and OSINT: Russia-Ukraine Conflict Case Studies}  



%

\author[label1,label2]{Sohail Ahmed Khan}[orcid=0000-0001-5351-2278]
\cormark[1]
\author[label3,label4]{Jan Gunnar Furuly}
\author[label3,label5]{Henrik Brattli Vold}
\author[label3,label6]{Rano Tahseen}
\author[label1,label2]{Duc-Tien Dang-Nguyen}[orcid=0000-0002-2761-2213]\cormark[1]

\affiliation[label1]{organization={University of Bergen},country={Norway}}
\affiliation[label2]{organization={MediaFutures},country={Norway}}
\affiliation[label3]{organization={Faktisk},
            country={Norway}}
\affiliation[label4]{organization={Aftenposten},
            country={Norway}}
\affiliation[label5]{organization={Institute of Journalism},country={Norway}}
\affiliation[label6]{organization={TV2},country={Norway}}




\cortext[1]{Corresponding authors.}



\begin{abstract}
This paper investigates the use of computational tools and Open-Source Intelligence (OSINT) techniques for verifying online multimedia content, with a specific focus on real-world cases from the Russia-Ukraine conflict. Over a nine-month period from April to December 2022, we examine verification workflows, tools, and case studies published by \faktiskbar. Our study showcases the effectiveness of diverse resources, including AI tools, geolocation tools, internet archives, and social media monitoring platforms, in enabling journalists and fact-checkers to efficiently process and corroborate evidence, ensuring the dissemination of accurate information. This research underscores the vital role of computational tools and OSINT techniques in promoting evidence-based reporting and combatting misinformation. We also touch on the current limitations of available tools and prospects for future developments in multimedia verification.
\end{abstract}




\maketitle

\section{Introduction}\label{sec:intro}

In today's digital era, social media platforms play a crucial role in news consumption and dissemination~\cite{pewrresearchreport, icfjreport2019}. They provide instant access to global news stories and a vast amount of real-time multimedia content. Journalists and news media organisations also actively use these platforms to stay updated on high-profile news events~\cite{Schifferesnewsverification, Niemanreports, icfjreport2019}.
The use of social media platforms in journalism however, acts as a double-edged sword. On the one hand, these platforms allow news media companies to reach wider audiences and stay connected with global news events. On the other hand, they are also used to spread mis/disinformation, especially during significant events like wars, elections, and pandemics~\cite{Ireton2018JournalismFN}. Consequently, it is crucial to responsibly utilise multimedia content from these platforms, particularly during breaking news events, by effectively fact-checking and verifying the information. Past incidents have shown instances where newsrooms and journalists failed to identify misleading content and shared it as reliable information~\cite{Thomson2020}. Such actions can have far-reaching consequences, eroding trust in news and damaging the reputation of journalists and media outlets~\cite{Ireton2018JournalismFN, Niemanreports}.

As a consequence, major news organisations such as The Associated Press\footnote{https://apnews.com/} and the British Broadcasting Corporation (BBC)\footnote{https://www.bbc.com/} have established dedicated teams specifically focused on verifying online multimedia content to ensure the accuracy and reliability of the information they disseminate further to their audiences. To aid in verification process, computational tools and Open-Source Intelligence (OSINT) techniques have become invaluable resources for the journalists who find themselves engaged with multimedia content on social media platforms on a daily basis. OSINT techniques involve gathering and analysing publicly available information from a wide range of online sources to verify the accuracy and credibility of claims~\cite{osint_definition}. A wide variety of computational tools can be employed to carry out OSINT analysis by collecting and processing content acquired from various online platforms such as social media websites, news articles, forums, online maps, and web archives~\cite{osint_team_tools}. 

In this paper, we examine how journalists and fact-checkers employ computational tools to analyse and debunk false information during the Russia-Ukraine conflict on social media. We focus on \faktiskbar%
\footnote{https://www.verifiserbar.no/}%
's work and the verified cases. We also discuss the tools they use, their needs, and challenges. Our objective is to map current practices and research, identify gaps, and propose improvements in fact-checking and journalism tools.
\begin{figure*}[ht]
\begin{minipage}[b]{1.0\linewidth}
  \centering
  \centerline{\includegraphics[width=1\linewidth]{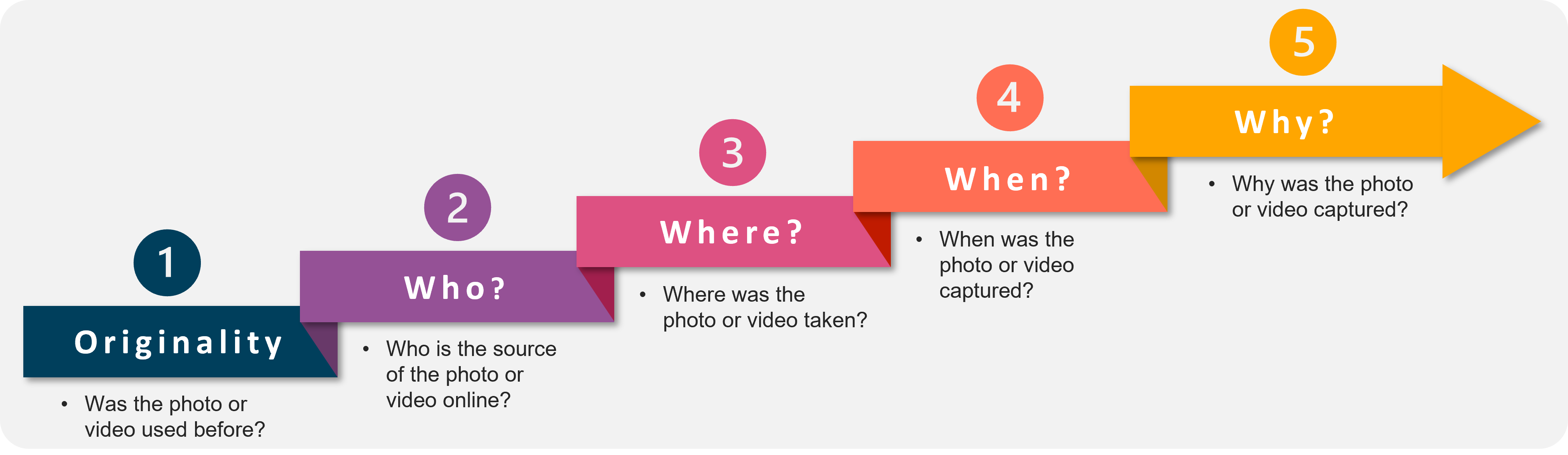}}
\end{minipage}
\caption{A common framework of multimedia verification followed by major news media outlets.
\faktiskbar also relied upon these basic steps in their workflows aimed at verifying multimedia content coming from Russia-Ukraine conflict. See Sections~\ref{subsec:5_core_element} and \ref{sec:verification} for more details.}
\label{fig:verificationworkflow}
\end{figure*}
In short, we present:
\begin{enumerate}[leftmargin=*,noitemsep,topsep=0pt]
    \item The verification workflows fact-checkers followed in verifying multimedia content; 
    \item Why and which computational tools they used and how they employed them in multimedia verification; 
    \item Real-world cases from the Russia-Ukraine conflict which were verified by \faktiskbar team;
    \item Some the limitations of current tools, and expectations for future computational tools in the context of multimedia verification.
\end{enumerate}

This paper is organised as follows: In Section~\ref{sec:section2}, we present the basic verification workflow and the specific guidelines at \faktiskbar to analyse multimedia content emerging in the context of the Russia-Ukraine conflict. Section~\ref{sec:verification} presents some of the cases verified by \faktiskbar and the tools they used. Section~\ref{sec:section4} offers a detailed overview of the analysis techniques and tools to analyse satellite images, along with some of the prominent cases they investigated. Finally, in Section~\ref{sec:vision}, we discuss the lessons learned, the gaps between practices and research and propose some improvements.

\begin{modernframe}
\textbf{Who are \faktisk and \faktiskbar}?

\faktisk\footnote{https://www.faktisk.no/}
is a fact-checking organisation that was created through a partnership between several major media companies in \anonycountry, including the state broadcasting company NRK\footnote{https://www.nrk.no/}, the media conglomerate Schibsted\footnote{https://schibsted.com/} and the liberal newspaper Dagbladet\footnote{https://www.dagbladet.no/}. \faktisk is a member of the International Fact-Checking Network and was established in 2017. In 2018, \faktisk announced a partnership with Facebook to fact-check content on the platform. 

In 2022, \faktisk established a special team called \faktiskbar. This team comprises of reporters, researchers and fact checkers from various media companies throughout Norway who work together to verify visual content and stories from the ongoing conflict in Ukraine. 
\end{modernframe}

\section{The Verification Workflow}
\label{sec:section2}

In the world of big news organisations, there is a widely-used workflow for checking online multimedia content. Bellingcat\footnote{https://www.bellingcat.com/}, a well-known group, follows a clear 5-step workflow of checking such content~\cite{bellingcat2021verificationguide}. This 5-step method is quite common and is also used by many other groups~\cite{Silverman2013VerificationH, khan2023visual}.
During the Russia-Ukraine conflict, \faktiskbar used the same 5-step workflow to verify the visual content and stories related to the Russia-Ukraine conflict. The workflow is summarised in Figure~\ref{fig:verificationworkflow} and in the next section, we present the five core elements of it.

\subsection{Five core elements of multimedia verification}
\label{subsec:5_core_element}

\begin{itemize}[noitemsep, leftmargin=*, wide=0pt]

\item \textbf{Originality.} During important news events, a multitude of images and videos flood the internet, encompassing both authentic and deceptive content. This includes instances where media may have been repurposed as part of ongoing internet trends or employed in coordinated campaigns on social media platforms. Consequently, if the originality of the content is not examined, subsequent steps of verification, such as identifying the source, location, time, and motivation, may be compromised, jeopardising the integrity of the verification process. Therefore, the initial and crucial step is to establish originality of the content item, ensuring that it has not been previously shared in a different context. This process of confirming originality safeguards the subsequent verification procedures and strengthens the overall process.
    
\item \textbf{Who?} The process of verifying the source (or who is behind this image/video) involves identifying the individual or group responsible for creating or capturing the original content. Through a comprehensive investigation of the source, including the content they shared in the past, and their social media footprint, valuable insights can be gained regarding the authenticity and credibility of the multimedia content being verified. Moreover, verifying the source is instrumental in detecting instances where the content has been misrepresented, repurposed, or taken out of context. By conducting thorough source verification, the accuracy and reliability of the multimedia content can be ensured.

\item \textbf{Where?} This component focuses on determining the geographical location where the content was captured or the location which is being shown in the content. Geolocation plays a pivotal role in the verification efforts undertaken by \faktiskbar as it aims to accurately determine the precise location of the depicted event or object showcased in the image or video. The establishment of location holds significant importance in evaluating the authenticity and context of the content. Through careful analysis of visual cues, landmarks, and distinctive elements within the multimedia content, investigators endeavour to geolocate the content and ascertain its alignment with the claimed location. This process often entails utilising geolocation tools, satellite imagery, and mapping data to cross-reference the visual information presented in the content. By verifying the "Where" component, instances of potential misattribution or false association of the content with a specific location can be identified.

\item \textbf{When?} In multimedia content verification, determining the precise time-frame (also Chronolocation) or temporal context in which the content was captured or occurred is also extremely important. Establishing the accurate timeline is essential in assessing the authenticity and relevance of the content item. By analysing various indicators such as metadata information, or contextual clues within the image/video itself such as, shadows, weather, time of the day etc, journalists can attempt to ascertain the specific moment or period when the event took place, or the content was captured. This verification process helps in detecting instances of potential misrepresentation, manipulation, or outdated information. 

\item \textbf{Why?} Finding out “Why” a certain image/video was shared by someone is almost impossible, since the act of sharing something online is driven by diverse motivations. While some individuals genuinely share content without having any bad purpose, others may have ulterior motives, such as advancing a particular political or personal agenda. Understanding the underlying motivation behind online posts is crucial for effectively assessing their credibility. When encountering a post from an individual known for disseminating misinformation, conspiracy theories, or biased content, it is prudent to exercise caution and conduct extensive verification to ensure the accuracy and reliability of the shared information. 
    
\end{itemize}

\begin{figure*}[ht!]
\centering
\includegraphics[width=0.9\linewidth]{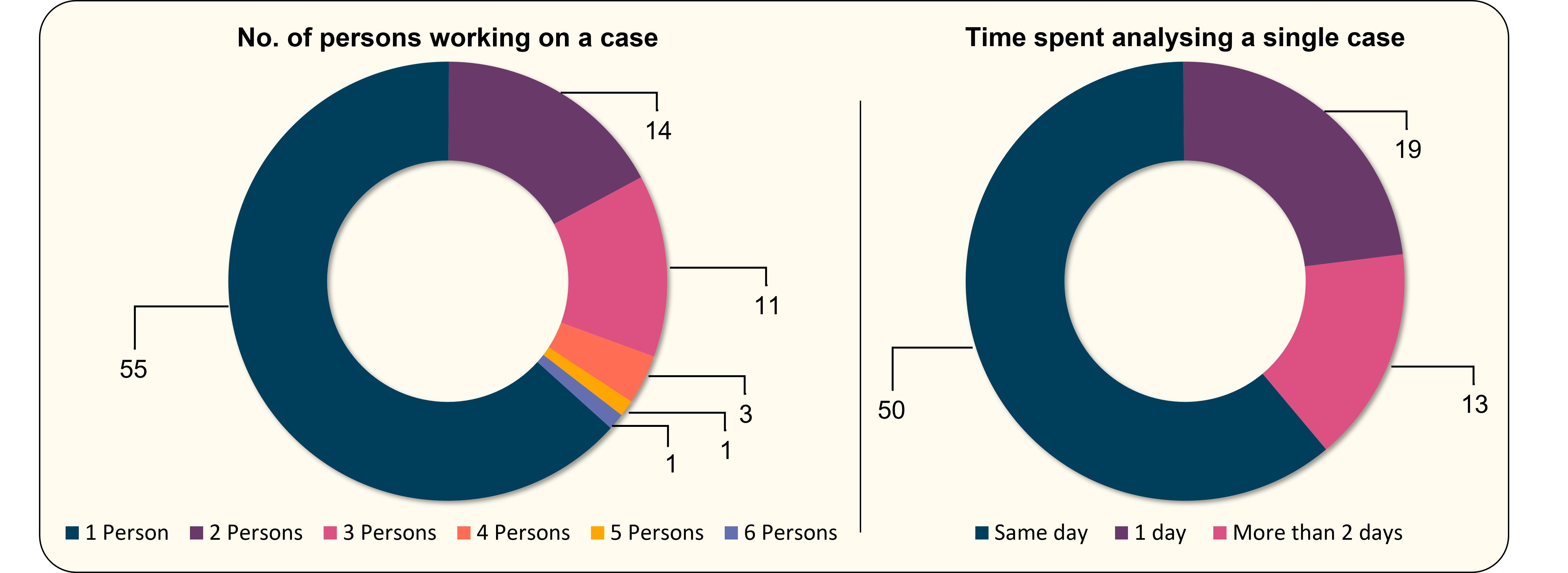}
\caption{On the left, we present number of distinct cases along with the number of persons working on a single case. On the right, in similar way we show the number of day(s) spent on solving each distinct case.}
\label{fig:persons_time}
\end{figure*}

\begin{figure*}[ht!]
\centering
\includegraphics[width=0.9\linewidth]{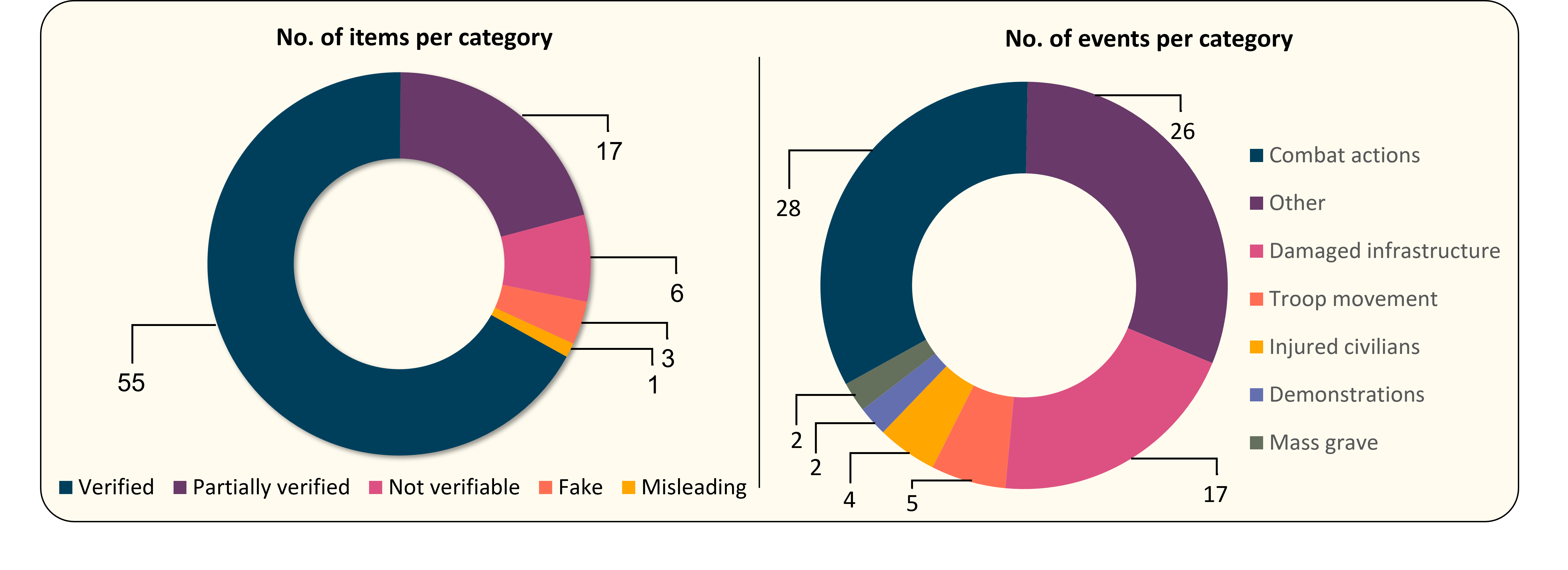}
\caption{On the left, we show the number of distinct cases belonging to each of the predefined categories. On the right, in similar fashion we show the number of cases belonging to each of the eight predefined event categories.}
\label{fig:categories}
\end{figure*}

\subsection{\faktisk specific guidelines}
In addition to the five mentioned elements, \faktiskbar has also incorporated some internal steps during and after verification has completed. We present these additional steps in this section below.

\subsubsection{The \faktiskbar verification database}
\label{sec:report_fields}

\faktiskbar has devised a database comprises from comprehensive reports and is limited for internal use only. All events to be verified are added to this verification database. In this study, we report in total 81 events from this database, summarised in Figures~\ref{fig:persons_time} and \ref{fig:categories}. The database contains these information: 
\begin{itemize}[leftmargin=*, noitemsep, wide=0pt]
    \item \textbf{ID:} The unique ID given to each case being verified.
    \item \textbf{Date of the Event:} The date of the occurrence of the event.
    \item \textbf{Date Published:} The date on which the image/video relating to the event was shared online.
    \item \textbf{Published:} Date on which the case was published to Mediabank\footnote{Mediabank is a service by \* accessible to journalists in \anonycountry. The Mediabank contains photos, videos, graphics related to news events.}. Journalists working at \faktisk publish a written report and any related material to the Mediabank associated to the case after completing the verification process.
    \item \textbf{Mediabank Link:} Link to the Mediabank case.
    \item \textbf{Job Status:} Current status of the job, includes (a) Completed, (b) Work in process, (c) On hold, (d) For control, and (e) Not a priority.
    \item \textbf{Status:} The classification assigned to the case, includes (a) Verified, (b) Not verifiable, (c) Partially verified, (d) False, and (e) Misleading.
    \item \textbf{Verified By:} The person/s who worked on the case.
    \item \textbf{Controlled By:} Peer control, i.e., a different person checks if the case has been dealt with in right manner, and does not lack any further verification.
    \item \textbf{Coordinates:} Coordinates of the location presented in the video/photo being verified.
    \item \textbf{Place (City, Area, Country):} The location where the event took place.
    \item \textbf{Category:} A specific category label is associated to each case being verified. The category is chosen according to the content being presented in the image/video. There are nine different predefined categories, (1) Injured civilians, (2) Troop transfer, (3) Combat actions, (4) Demonstrations, (5) Damaged infrastructure, (6) Hope and heart, (7) Mass grave, (8) Still digging, i.e., gathering more information about the content, and (9) Other. Figure~\ref{fig:categories} summarises how the categories distributed.
    \item \textbf{Level of Violence:} Has five different subcategories, (1) None - Military presence, (2) Mild - Explosion/ Destruction/ Mild damage, (3) Moderate - Bombing/Moderate damage, (4) Heavy - Serious damage/Slapped corpse, and (5) Very strong - Dead or injured bodies/Explicitly gross or particularly close violence.  
    \item \textbf{Sender:} (1) Pro-Russia, (2) Pro-Ukrainian, (3) Neutral, and (4) Other.
    \item \textbf{Description:} A short summary of what is being presented in the image/video.
    \item \textbf{Content Link:} Link to the originally shared image/video, e.g., link to the Tweet or Facebook post.
    \item \textbf{Video or Photo Link:} Direct link to the image/video after being saved in the private \faktisk Google drive folder.
    \item \textbf{Map Link:} Link to the private Google drive work folder.
    \item \textbf{Comments:} Relevant comments directed towards other \faktisk colleagues.
\end{itemize}

\subsubsection{The Data Collection and Publishing Process}
The \faktiskbar team receives verification requests through a public email tip line, ensuring open channels for information submission. Additionally, they receive cases directly from parent media companies 
(VG\footnote{https://www.vg.no/}, Dagbladet, NRK, TV 2\footnote{https://www.tv2.no/}, Amedia\footnote{https://www.amedia.no/}).
They also monitor social media on platforms like Twitter, Facebook, Telegram, Reddit, using tools like Dataminr\footnote{https://www.dataminr.com/} and Tweetdeck\footnote{https://tweetdeck.twitter.com/}. They follow Telegram channels (e.g., Telehunt, TelegramExplore, RussianDisinformation Dashboard, Amplify Ukraine), Twitter accounts (e.g., Geoverified), and Discord servers (e.g., Bellingcat Discord server). Additionally, they use resources like the Monitor Map\footnote{https://maphub.net/Cen4infoRes/russian-ukraine-monitor} and LiveuaMap\footnote{https://liveuamap.com/} to gather real-time information and track the situation on the ground.

When a relevant event/case is identified for verification, it is logged in a dedicated entry in the dataset, as mentioned earlier. In addition, a separate Google Drive folder is created with a unique identifier. This folder houses various resources, such as raw files (images/videos), screenshots, and geolocation details (coordinates). Furthermore, a verification document is generated within the project folder using a base template. The template comprises several main sections, including \textit{``The public report text"}, \textit{``What does the video/picture actually show?"}, \textit{``Publishing context"}, \textit{``Geolocation"}, and \textit{``Time verification"} Each section provides a brief explanation of the verification process, outlining the steps taken, tools utilised, and other relevant information.
 
Once the verification document is complete, the case is updated to \textit{``ready for inspection"} for a peer control (buddy check), a standard \faktiskbar procedure. This step ensures a second opinion on verification, confirms document accuracy, and involves another person thoroughly reviewing and approving the document. After a positive evaluation, the document can be published with related media through Mediebank. An email notifies the tipper, typically a journalist in \anonycountry, and if the event is newsworthy, another email goes to a predefined list of media personnel. Sometimes, a full news story is written and published on the \faktiskbar platform, especially when the event gains attention among their journalists.

\section{Verifying Russia-Ukraine Conflict}
\label{sec:verification}
This section provides an overview of \faktiskbar's process for conducting visual multimedia content verification in relation to the Ukraine conflict, employing the five verification elements presented earlier. 

\subsection{Establishing originality}


There are various methods available to authenticate an image or video. Among them, reverse image search is widely utilised. This technique involves uploading the image (or a single frame in the case of a video) to search engines like Google, Bing, or Yandex\footnote{https://yandex.com/}, or using dedicated tools like TinEye\footnote{https://tineye.com/} or Google Lens\footnote{https://lens.google/} to search for similar instances of the image across the web. For video verification, the InVid-WeVerify verification plugin\footnote{https://weverify.eu/verification-plugin/}~\cite{MezarisInVid} is commonly employed. This plugin aids in extracting individual frames from videos to conduct reverse image searches. Additionally, software programs such as VLC\footnote{https://www.videolan.org/} or Adobe Premiere\footnote{https://www.adobe.com/products/premiere.html} are sometimes employed to extract keyframes from videos. By performing a reverse image search, it becomes possible to uncover any alterations or contextual misuse of the image or video. If a previous version of the same image or video is discovered online, it strongly suggests recycling and out-of-context presentation.

Another method of verifying the authenticity of visual content is through the analysis of metadata associated with an image or video. Metadata contains valuable information about the creation time and location of the content, which can aid in assessing its authenticity. However, it is important to note that the metadata is often absent in visual content downloaded from social media platforms. This is because these platforms typically strip the metadata during the upload process to save storage space~\cite{Pasquini2021}. 
Also, caution should be exercised when dealing with metadata information, as it can be subject to modification or manipulation.

In addition to the above methods, the embedded watermarks or logos found in certain images and videos can sometimes be instrumental in determining the content's source and providing valuable leads for further investigation. Also, journalists carry out keyword searches across various social media platforms, such as Telegram channels, Twitter, and Discord servers, using queries related to the events depicted in the image or video. These approaches helps them gather additional information about the content's origin. Facial recognition software, such as PimEyes\footnote{https://pimeyes.com/en} and Search4faces\footnote{https://search4faces.com/}, are also occasionally employed when images or videos contain identifiable faces, aiding in source identification.

\begin{figure*}[t!]
  \centering
  \centerline{\includegraphics[width=0.85\linewidth]{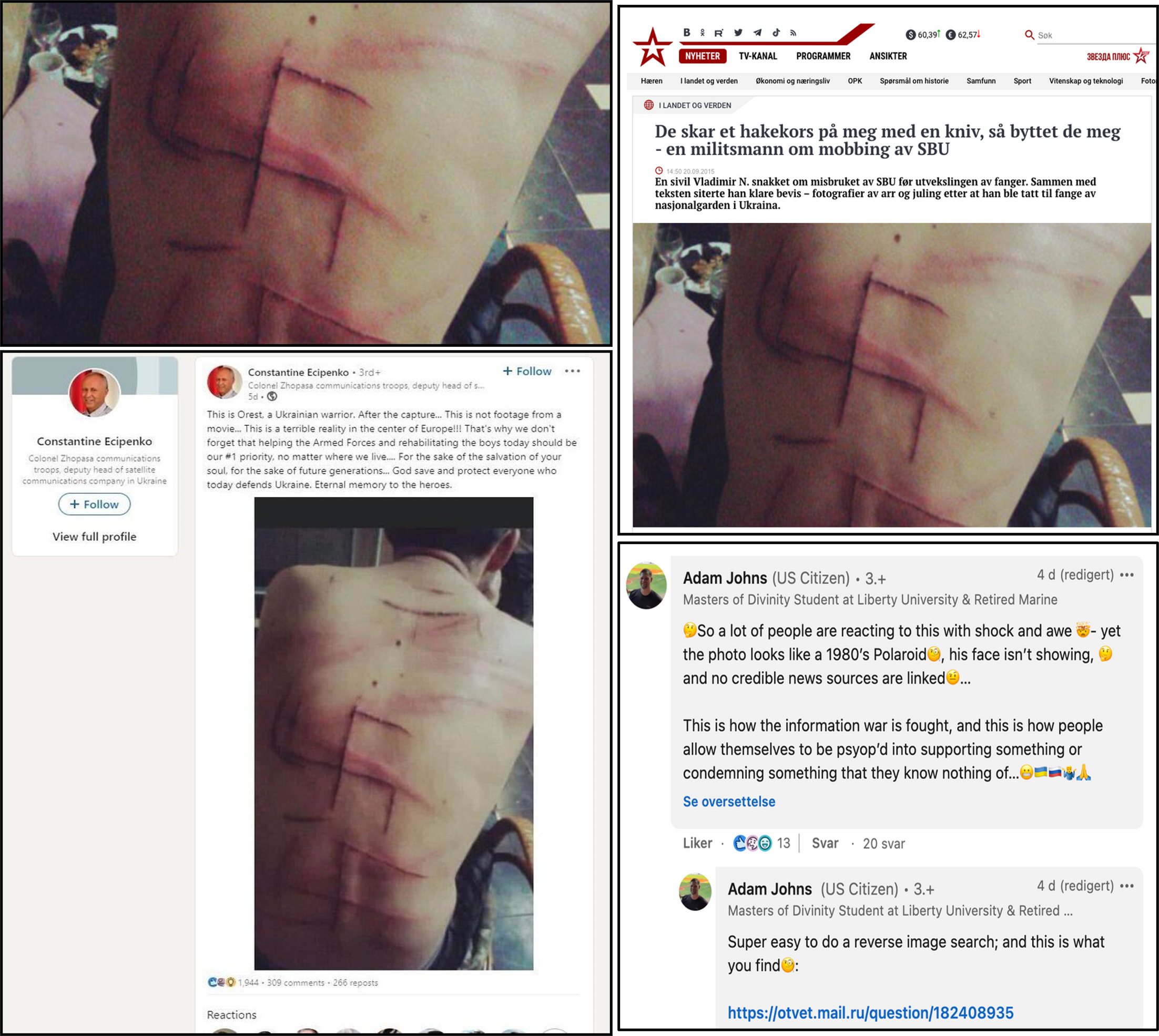}}
  \caption{Misused image during Russia-Ukraine conflict: alleged Ukrainian soldier with swastika carved, proven unrelated through reverse image search. PHOTOS: \faktiskbar.}
\label{fig:reverseimagesearch}
\end{figure*}

An instance highlighting the misuse of old images to propagate misinformation during the Russia-Ukraine conflict was the case of an alleged Ukrainian soldier named Orest, who was depicted in a viral post with a swastika carved into his back by Russian soldiers. This image resurfaced on social media platforms, circulated by various users, including a retired Norwegian military personnel and a well-known Ukraine war specialist, Retired Lieutenant General Arne Bård Dalshaug, who often features in Norwegian media. To verify the authenticity of the image, \faktiskbar conducted a simple reverse image search, revealing that the same image had been previously employed in unrelated contexts. Figure~\ref{fig:reverseimagesearch} provides a visual representation of this case.

\begin{figure*}[t!]
  \centering
  \centerline{\includegraphics[width=0.8\linewidth]{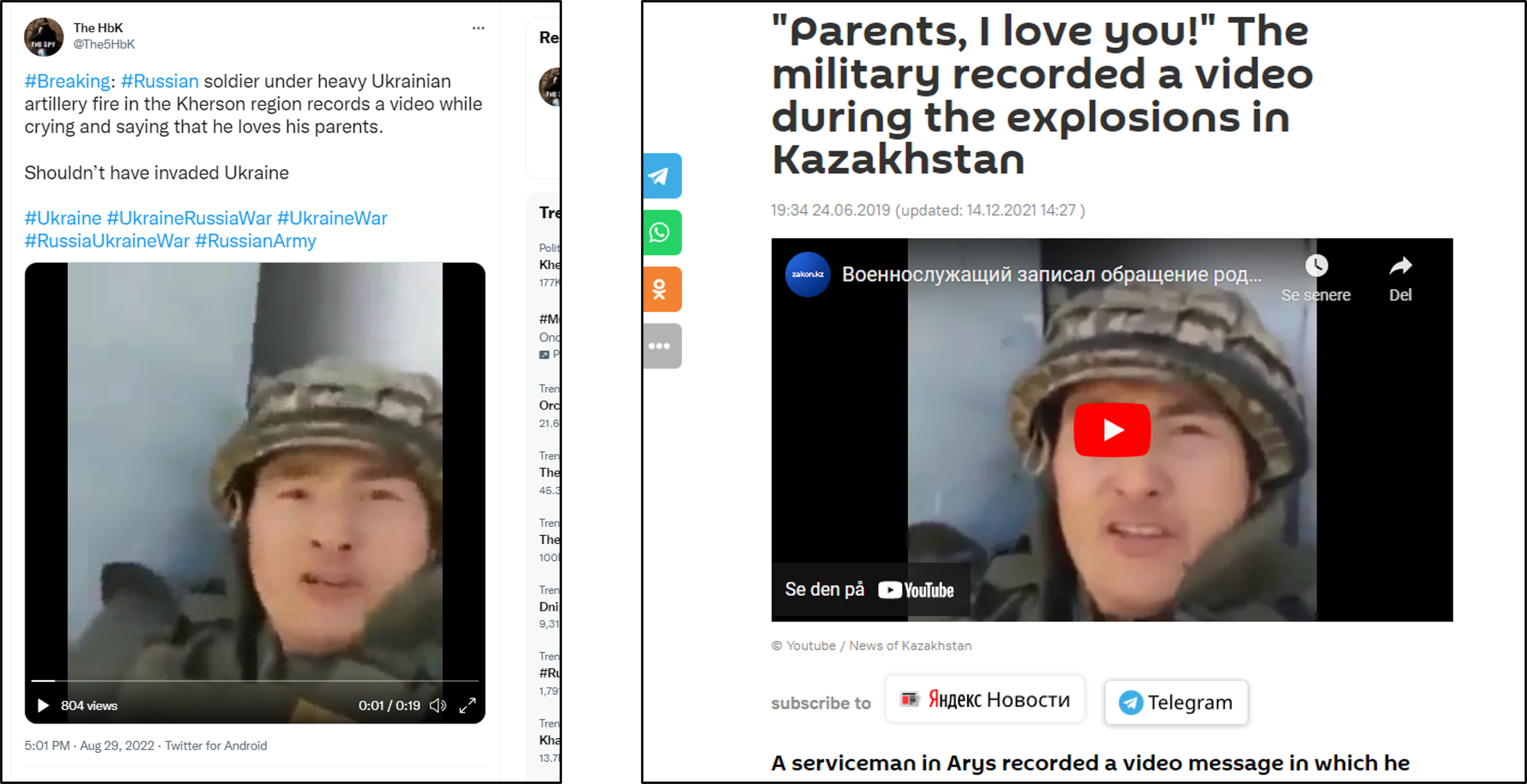}}
\caption{Photo on the left shows a soldier crying in front of the camera along with a misleading caption relating to Ukrainian counteroffensive in the Kherson region against Russian forces. Photo on the right shows the original video posted on a Kazakhstani TV channel, KTK, showing a soldier crying and saying to the camera, "If I don't survive, parents, I love you". PHOTOS: \faktiskbar.}
\label{fig:soldier_ooc}
\end{figure*}

\faktiskbar investigated another case involving miscontextualised content shared on social media, purportedly related to the conflict in Ukraine. On August 29, 2022, a video emerged claiming to depict a Russian soldier in distress. The footage showed a soldier crying on camera while the sound of artillery echoed in the background. The accompanying caption claimed that the soldier was Russian and under intense fire in the Kherson region, during Ukraine's counteroffensive against Russian forces at that time. Refer to the left photo in Figure~\ref{fig:soldier_ooc} for visual reference.

Through the use of the facial recognition tool PimEyes\footnote{https://pimeyes.com} and conducting reverse image searches, \faktiskbar discovered that the same video had been previously published in 2019, including on the Kazakhstani TV channel ``KTK" and the Russian state-owned media outlet ``Sputnik". According to KTK, the soldier in the video was heard saying, \textit{``If I don't survive, parents, I love you"}. For visual reference, please see the photo on the right side of Figure~\ref{fig:soldier_ooc}. Additionally, \faktiskbar utilised the PimEyes facial recognition software to investigate and identify certain individuals from Azovstal and the notorious Wagner group.

\subsection{Who is behind it?}
Another crucial element of multimedia verification is to ascertain the identity of the original creator, uploader, or sharer of the content, and if possible establish communication with them. Journalists pose pertinent questions to the individuals, such as the location where they were present when capturing the footage, their observations of the incident, and the type of device used to record the content. Interviewing the person/entity help uncover any potential false information being disseminated, whether intentional or unintentional. By asking direct questions, journalists often prompt the person to admit if they did not actually film the footage themselves. Moreover, answers to these questions can be cross-referenced with available information, such as examining the Exif data in a photo or comparing the video with Google Street View, as detailed in subsequent sections. In case the photo/video is uploaded to a social media platform, metadata information is often lost, but journalists can also request original file, or supplementary corroborating evidence, such as additional images or footage, to verify the individual's presence at the location.

In case its not possible to establish contact with the person/entity who uploaded or shared the content online, journalists utilise alternative methods to authenticate the content. One method is to conduct a social media analysis by searching for the user (by using their usernames, other relevant information) who shared the content on other prominent social media platforms, online forums, and person search engines such as PeopleFinder.com\footnote{https://www.peoplefinder.com/}, Spokeo.com\footnote{https://www.spokeo.com/}, Webmii.com\footnote{https://webmii.com/}, or Pipl.com\footnote{https://pipl.com/}. In case a website is under investigation, online tools such as Who.is\footnote{https://who.is/}, DNSChecker.org\footnote{https://dnschecker.org/} and other similar domain search engines can be employed. The other solution is to analyse any associated metadata information of the content in case it is available. There are numerous metadata extraction and analysis tools available online such as, exifdata.com\footnote{https://exifdata.com/}, Forensically\footnote{https://29a.ch/photo-forensics/}, FotoForensics\footnote{https://fotoforensics.com/} or FotoVerifier\footnote{https://dedigi.fotoverifier.eu/}~\cite{tran2022dedigi}.

This endeavour can yield valuable insights into the user's online presence, activities, and behavioural patterns. By conducting a thorough analysis of their interactions, posts, and engagement with others, journalists can develop a clearer understanding of the user's credibility, consistency, and potential motivations. Moreover, exploring relevant discussions or mentions of the content on these platforms can contribute to the verification process by unearthing additional perspectives or corroborating information from other sources.

Additionally, facial recognition software such as PimEyes can be employed to aid in authentication. By using the photo of the uploader, this software can conduct a search across the internet to identify other instances where the same face appears. This can help verify if the person behind the content has a consistent online presence or if their identity has been associated with other activities or accounts. 

\subsection{Where is this/What place is being showed in the image/video?}

\begin{figure*}[t!]
  \centering
  \includegraphics[width=\linewidth]{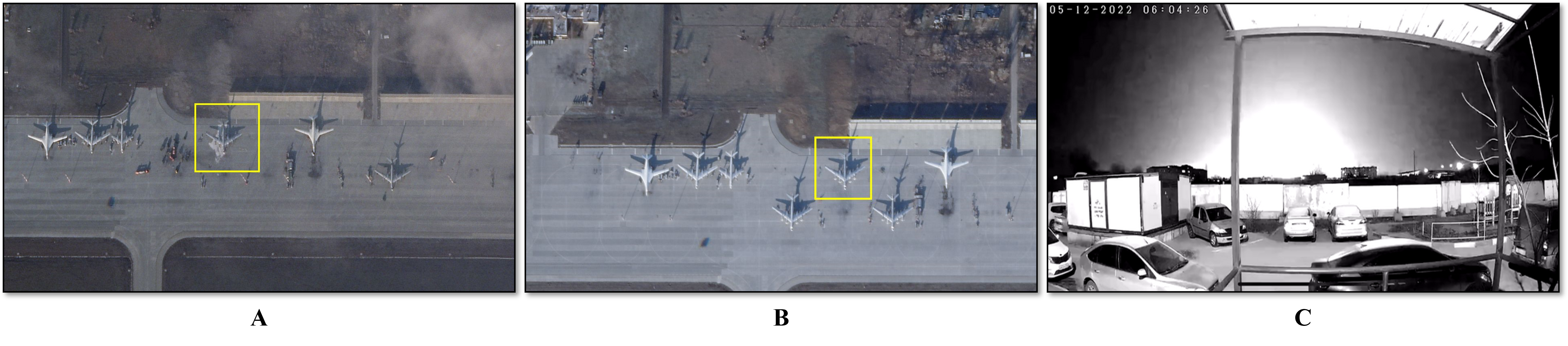}
\caption{On the night of Monday, December 5, 2022, a powerful explosion occurred at Engels airfield, located 720 kilometers southeast of Moscow. Image \emph{(A)} shows a TU-95 bomber situated on the ground amidst fire-retardant foam, with two fire trucks and personnel present near the affected aircraft. This photograph, captured on Tuesday, December 6, 2022, was taken using Planet Labs. Image \emph{(B)} showcases a pre-attack scene captured by Planet Labs on Sunday morning, December 4, 2022, revealing the plane in the same position. No apparent irregularities were observed on this satellite image from Sunday morning. Image \emph{(C)} is a still frame extracted from a surveillance camera video, obtained from Telegram, recorded approximately 3 kilometers away from the airbase. PHOTOS: TELEGRAM, PLANET LABS, \faktiskbar.}
\label{fig:engels}
\end{figure*}

\faktiskbar utilise various robust online mapping tools to geolocate events depicted in images and videos related to the Russia-Ukraine conflict. Among the prominent geolocation tools they rely on are Google Maps, Google Earth Pro, Yandex Maps\footnote{https://yandex.com/maps/}, Planet Labs, and NASA FIRMS\footnote{https://firms.modaps.eosdis.nasa.gov/}. Here, we illustrate the case of an explosion at Engels airbase in Russia, which took place on December 5, 2022, using satellite imagery obtained from Planet Labs.

Reports on social media platforms, including Telegram and Twitter, indicated a drone attack explosion near Russia's Engels airbase in the Saratov region on December 5, 2022, about 74 kilometers from the Ukrainian border. Two CCTV videos from nearby apartments captured the explosion, although the exact location and direct link to the airbase remained unconfirmed at the time.

To investigate, \faktiskbar started to gather evidence relating to the incident. They obtained recent satellite images of the airbase from Planet Labs, specifically focusing on the week prior to the alleged incident. The analysis of these images revealed the presence of strategic bombers, including Tu-160 and Tu-95 aircraft, stationed at the airbase. Notably, there were no visible indications of damage to either the airbase or the aircraft. Figure~\ref{fig:engels}~(B) depicts the image of the airbase captured on Sunday, December 4, 2022, a day before the explosion was reportedly witnessed.
However, an image captured by Planet Labs on Tuesday, December 6, at 08:00 Norwegian time, revealed a significant development at Engels base. The satellite image showcased a TU-95 aircraft with propellers, surrounded by fire-retardant foam, and accompanied by several personnel and fire trucks, Figure~\ref{fig:engels}~(A). 

By leveraging multiple online mapping tools, \faktiskbar effectively conducted geolocation analysis of the aforementioned videos capturing the explosion from a distance. The two videos recorded by surveillance cameras and shared on Telegram relating to the explosion provided additional important evidence. Figure~\ref{fig:engels}(C) showcases a snapshot extracted from one of the CCTV footage. The locations depicted in the videos were cross-referenced with Google Street View to validate their accuracy. Furthermore, employing sound speed calculations, it was confirmed that these locations corresponded to the proximity of the airport. Remarkably, the direction of the explosion captured in the CCTV footage, as depicted in Figure~\ref{fig:engels}(C), aligns with the area of the airbase. With the combined information gathered, it is evident that an explosion indeed took place at the Engels airbase, potentially causing damage to an aircraft, as depicted in Figure~\ref{fig:engels}~(A). \faktiskbar made international headlines as the first newsroom to report on this incident, providing an in-depth account of damage inflicted by the drone attack at the Engels airbase.

\begin{figure*}[t]
\hspace*{-0.1cm}
\begin{minipage}[b]{1.0\linewidth}
  \centering
  \centerline{\includegraphics[width=1\linewidth, height=0.35\linewidth]{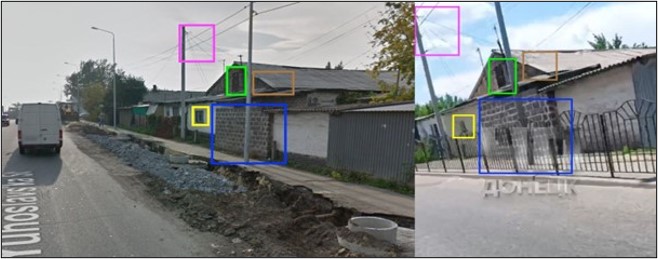}}
\end{minipage}
\caption{The ``Donetsk Car Wash" case. Image on the right is from the footage, whereas image on the left is from Google Street View. The coloured boxes highlight the matching house facade, pole, and the fence.}
\label{fig:carwash}
\end{figure*}

In the ``Donetsk Car Wash" case, readily available tools like Google Maps, VLC, Google Translate, and Yandex Maps played a vital role in verifying an online video shared on social media. The video, shot from a moving vehicle, showed an artillery shell on a road in Donetsk. \faktiskbar used VLC Media Player to navigate through the frames, focusing on distinct features like a house facade and a car wash sign. Google Translate helped translate the Ukrainian term for ``car wash," and Yandex Maps provided potential locations in Donetsk, later confirmed through Google Maps and Yandex Maps, using Street View to match the house facades (Figure~\ref{fig:carwash}). Note that this task was challenging due to outdated Yandex images from 2010, while Google Street View images from 2011 revealed key details about the location.

\subsection{When did this happen?}

The timestamps associated with content shared on social media platforms do not always indicate the original date of capture. It is also important to note that visuals can be shared multiple times across different platforms, each with its own distinct timestamp. This creates a challenge in accurately determining the true date of creation for a visual multimedia content item. Journalists are aware of this issue and, to address it, they typically include the date and time of capture when publishing such content. This helps provide a more reliable reference for understanding the chronology of events. 
However, ordinary people often fail to provide the precise date of capture for the content they share. Even if they do mention a date, it is important to approach such information with caution due to the abundance of misinformation present on these platforms. It is thus crucial for the journalists to further investigate, and find the actual date of capture.

Visual content, such as images and videos, contains embedded metadata known as Exif headers. These headers hold valuable information about the date, time, device, camera settings, and even GPS location of the captured content. However, when content is downloaded from social media platforms, the Exif headers are often removed to save storage space~\cite{Pasquini2021}. Consequently, verifying the authenticity of Exif information becomes challenging. To overcome this limitation, journalists may request the original image or footage directly from the eyewitness or the individual who uploaded and shared the content~\cite{Silverman2013VerificationH}. By obtaining the unaltered file, the Exif data can be examined to verify its accuracy and establish a more reliable timeline. 
However, it is important to exercise caution when relying on the information found in the Exif header, as it can be easily modified and should be approached with care.

When visual content lacks accompanying metadata or visible clues to determine the date and time of an event, journalists at \faktiskbar employ alternative techniques such as shadow analysis, and historical weather pattern analysis. These methods allow them to make approximate estimations regarding the date and/or time depicted in the visuals. For details see Table~\ref{tab:when}. 
The following case exemplifies how \faktiskbar utilised these approaches to infer the dates by examining visual details and studying historical weather patterns.

\begin{figure*}[t!]
  \centering
  \centerline{\includegraphics[width=\linewidth]{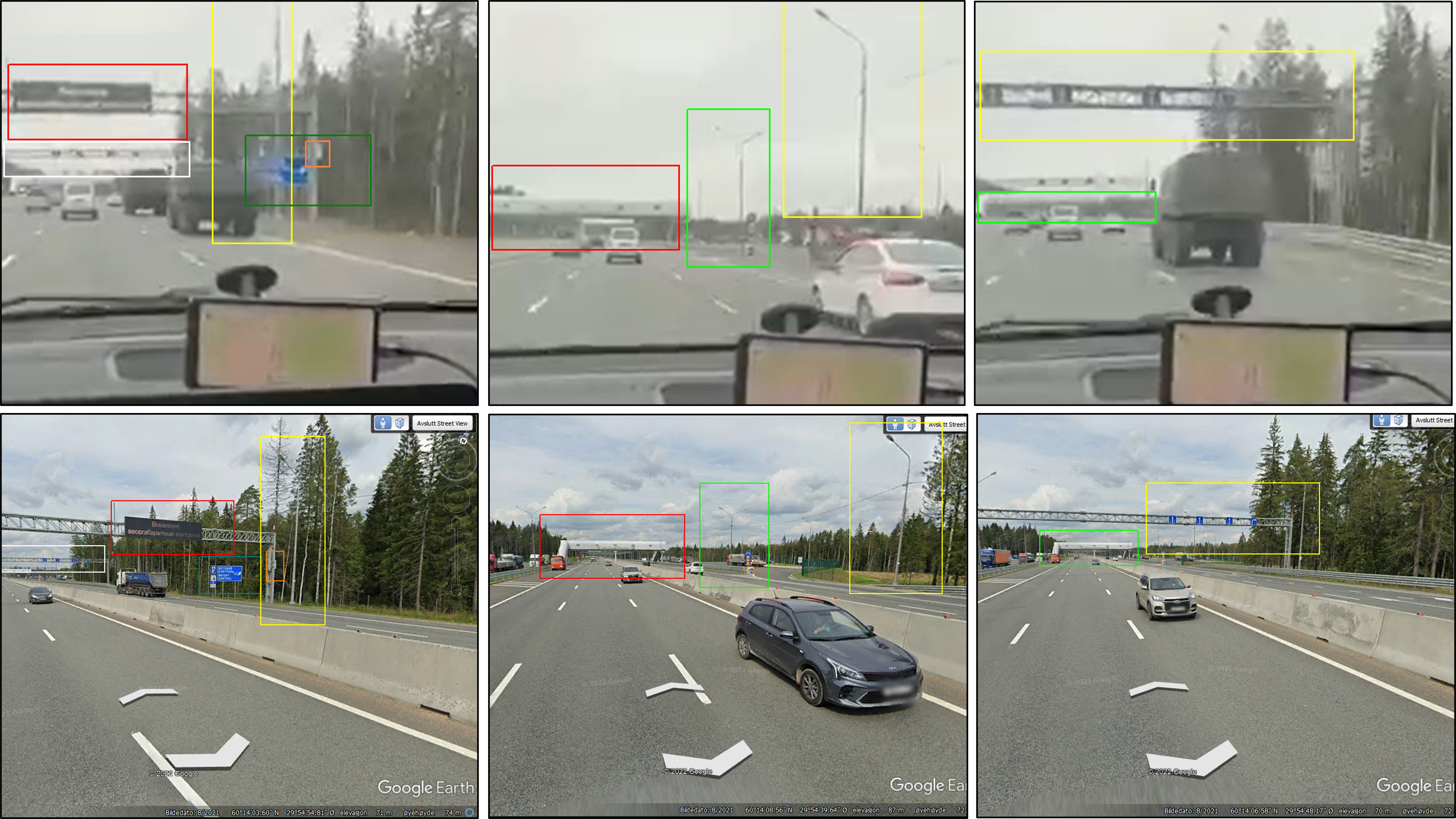}}
\caption{Photos on the top row shows frames from the video, whereas photos on the bottom show Google Street View images. The colored squares compare the landmarks present on the highway correlating photos from the video with that of the photos from Google Street View. PHOTOS: \faktiskbar.}
\label{fig:iskander}
\end{figure*}

In the middle of May 2022, \faktiskbar received a report regarding a video that claimed Russia was transporting missile-equipped vehicles towards its border with Finland. The video, recorded from a car, depicted a convoy of military missile vehicles traveling on a highway. The accompanying caption suggested that this activity was a response to Finland's pursuit of NATO membership. To verify the video's authenticity and determine its location, \faktiskbar engaged in a comprehensive investigation.
By carefully examining the visual details within the video and analysing the Russian highway network, \faktiskbar successfully geolocated the exact filming site. This was achieved by identifying specific road signs visible in the footage (see Figure~\ref{fig:iskander}). The confirmed location was found to be approximately an hour and fifteen minutes' drive from Vyborg, situated 127 kilometers away from the Finnish border. Through these verification efforts, \faktiskbar shed some light on the context and accuracy of the video in question.

\begin{figure*}[t!]
  \centering
  \centerline{\includegraphics[width=0.8\linewidth]{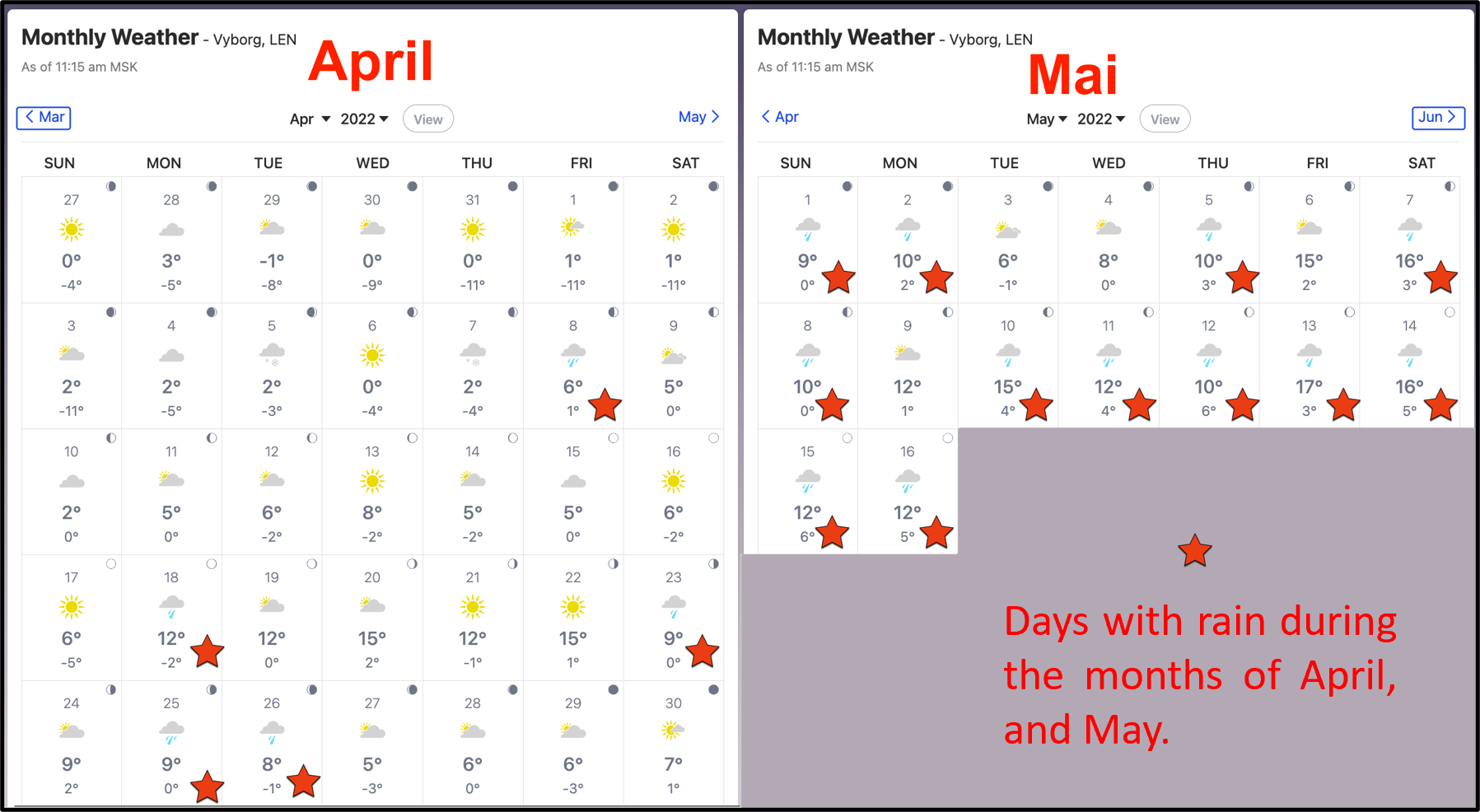}}
\caption{Photos show snapshots from the website \textit{Weather.com}. The red stars represent the days with rain during the months of April, and May 2022. By using this weather data, \faktiskbar narrowed down the possible candidate days on which the video relating to the missile movement towards the Finnish border was recorded. PHOTOS: \faktiskbar.}
\label{fig:iskander_weather}
\end{figure*}

However, determining whether the video was recorded after the initiation of the NATO process posed another challenge. To address this, \faktiskbar leveraged historical weather patterns as a valuable tool in establishing the possible time-frame for the video's filming. By examining satellite imagery from the Sentinel Hub\footnote{https://www.sentinel-hub.com/}, it was confirmed that snow was present in the area as recently as April 23, 2022. Since the video displayed no signs of snow, this information played a crucial role in narrowing down the potential filming period.
Notably, the video depicted the car's windshield wipers in operation, indicating rainfall on the day of recording. To validate this, \faktiskbar consulted historical weather data from Weather.com\footnote{https://weather.com/} to identify the days when rain was reported in the vicinity. Although it was not possible to ascertain with absolute certainty that the clip was from May 16, the combination of historical weather data and satellite imagery from the Sentinel Hub strongly suggested that the video was captured in May, rather than April—the same month in which Finland commenced the NATO accession process. This verification process became the primary news story across the Nordics on May 17.

\begin{figure*}[t]
\hspace*{-0.1cm}
\begin{minipage}[b]{1.0\linewidth}
  \centering
  \centerline{\includegraphics[width=1\linewidth]{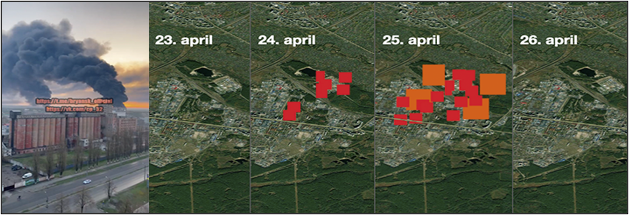}}
\end{minipage}
\caption{This figure shows the on the left a frame from the video showing a burning fuel depot. The images on the right are taken from NASA FIRMS on 4 different dates. The NASA FIRM images show high temperature in the area where the fuel depot is located between 24th and 25th of April. Through this analysis the \faktiskbar established that the video showing burning fuel depot was indeed genuine and not manipulated.}
\label{fig:firemaps}
\end{figure*}

In yet another investigation related to the Ukraine conflict, \faktiskbar delved into a case involving a major fire in the Russian city of Bryansk. The incident gained attention as Russian state media alleged that Ukraine was responsible for attacking two fuel depots in the city. However, verifying the videos depicting the incident proved challenging due to poor lighting conditions. This situation held significant importance as it could have marked the first instance of Ukraine launching a counterattack on Russian territory during the conflict.
To shed light on the matter, \faktiskbar journalists employed online mapping tools such as Google Maps and NASA FIRMS. By utilising NASA's fire maps, the journalists were able to confirm the dates of the attacks. The fire maps showcased significantly higher heat signatures in the relevant areas, providing strong evidence for the occurrence of the incidents. For reference, see Figure~\ref{fig:firemaps}.

\subsection{Why is it shared online?}

Understanding the motive behind shared social media content is essential for journalists verifying its authenticity. This comprehension aids in evaluating reliability, detecting manipulation, and uncovering agenda-driven context. It also safeguards against personal biases, ensuring impartial and truthful reporting. Investigating content motives is integral to verification, promoting accuracy and ethical reporting.

We present a case investigated by \faktiskbar, focusing on a viral propaganda video that gained widespread attention on social media in early September 2022. The video claimed to be a production of Gazprom, a Russian energy company with both state-owned and private ownership. It showcased the icy conditions in Europe as a result of purported gas shutdown due to sanctions. The video garnered extensive sharing among Russian individuals and was subsequently picked up and shared by various Western social media accounts, amplifying its reach.
Upon investigation, \faktiskbar did not find any official videos shared by Gazprom that aligned with the described content. However, subsequent online evidence emerged suggesting that the viral video was not produced by Gazprom, but rather by a Russian journalist named Arthur Khodyrev. When questioned about the video (not by \faktiskbar), Khodyrev admitted to creating it on his own accord without any monetary compensation. His intention behind the video was to showcase to Europe the perceived repercussions of imposing sanctions on Russia and refraining from purchasing its gas, which he referred to as a form of ``gas suicide".
After conducting additional investigation into journalist Arthur Khodyrev, \faktiskbar uncovered several noteworthy details. Firstly, it was discovered that the video in question was composed of various pre-existing clips that were pieced together. Secondly, the inclusion of footage featuring the city of Krasnoyarsk in Siberia was deemed irrelevant as it is not linked to the Russian gas network. Lastly, the background music used in the video was of low quality, as it was extracted from a live recording of a song that had been uploaded to YouTube back in 2016.
By thoroughly examining all the available evidence, it becomes apparent that the main intention behind the video was to provoke unrest among the European population. The video aimed to instigate doubt and raise questions about the decisions made by their respective governments, particularly regarding the imposition of sanctions on Russia and the choice to refrain from purchasing Russian gas.

In the earlier case, \faktiskbar swiftly identified the motivation behind a specific piece of content without the need for an extensive investigation. However, it's crucial to recognise that in numerous instances, uncovering motivation may be more complex. The following steps can be followed to gather additional information and shed light on the underlying motivation:
\begin{itemize}[leftmargin=*, noitemsep, wide=0pt]
    \item \textbf{Source Verification:} To understand the motivation behind a shared multimedia content, it is vital to confirm the content's source, including the person or organisation responsible for sharing it online. This involves a thorough online background check. For content posted on websites, examining the website can offer valuable insights, such as through DNS analysis. Identifying the source helps uncover the motivations and intentions behind its internet dissemination.
    \item \textbf{Contextual Analysis:} Evaluate contextual elements such as captions, mentions, and group affiliations to gain insight into why the content was shared. This analysis helps reveal the intentions behind its dissemination and its impact on various groups.
    \item \textbf{Explore Related Posts:} Conduct a search for related posts or content from the same source or event to build a comprehensive understanding of the context. Examining additional material sheds light on patterns and the overarching narrative.
    \item \textbf{Direct Communication:} If uncertainty persists regarding the content's motivation, consider reaching out to the original poster. Initiating a conversation provides an opportunity to seek clarification, inquire about their intentions, and gain deeper insights into the content's context and purpose.
\end{itemize}

\section{Satellite Imagery Analysis}
\label{sec:section4}
Satellite imagery analysis plays a significant role in modern verification journalism~\cite{reuters-satellite}. By providing detailed and accurate pictures of the Earth's surface, satellite imagery can offer valuable insights that may have otherwise gone unnoticed.
Journalists/fact-checkers often use satellite imagery to verify or challenge claims made by individuals and/or governments, military forces, or other groups. For instance, if a nation rejects  that it has placed its forces in a particular area, satellite imagery can be used to verify the claims, i.e., whether there are troops or equipment present or not at a specific location. Similarly, satellite imagery can be employed to track the movement of military personnel/equipment and identify potential conflict areas.
Satellite imagery analysis can also be used to examine environmental issues such as deforestation, illegal mining, and oil spills. It can also be used to track the spread of infections and natural disasters, or to monitor changes in weather patterns.

During the coverage of the Russia-Ukraine conflict, \faktiskbar sought to acquire access to regularly updated satellite images of the conflict zone and key locations within Russia, Ukraine, and Belarus that played a crucial role in the conflict.
Together with free services, they also exploited Maxar\footnote{https://www.maxar.com/} and Planet Labs\footnote{https://www.planet.com/}, the two well-known providers of high-resolution satellite imagery, offering regular updates. 
%
%
In the upcoming sections, we highlight noteworthy cases where \faktiskbar leveraged their partnership with Planet Labs to analyse and examine high-resolution satellite imagery, shedding light on critical developments.

\subsection{The Strategic Bombers Case}
\begin{figure*}[t!]
  \centering
  \includegraphics[width=\linewidth]{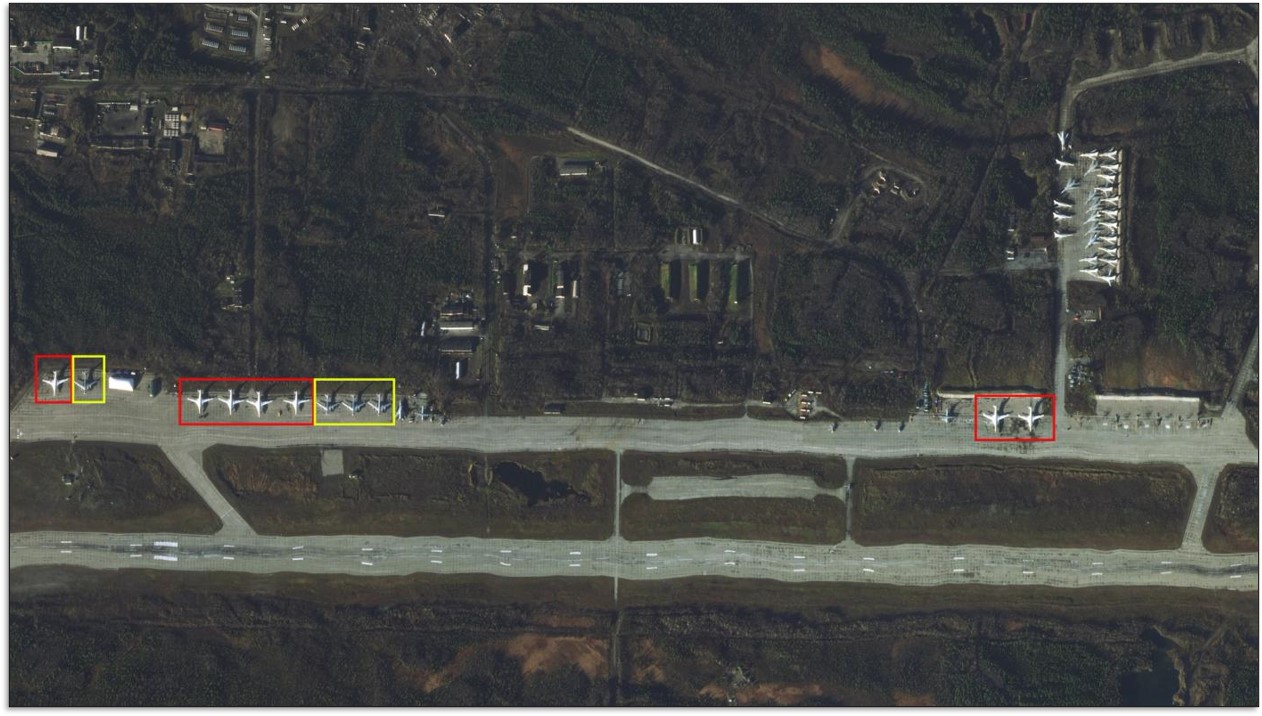}
\caption{A satellite image taken on 7 October shows seven Tu-160 strategic bombers (marked in red) and four Tu-95 aircraft (marked in yellow) at the Olenja airbase on the Kola Peninsula. PHOTO: Planet Labs and \faktiskbar.}
\label{fig:bombersatborders}
\includegraphics[width=0.6\linewidth]{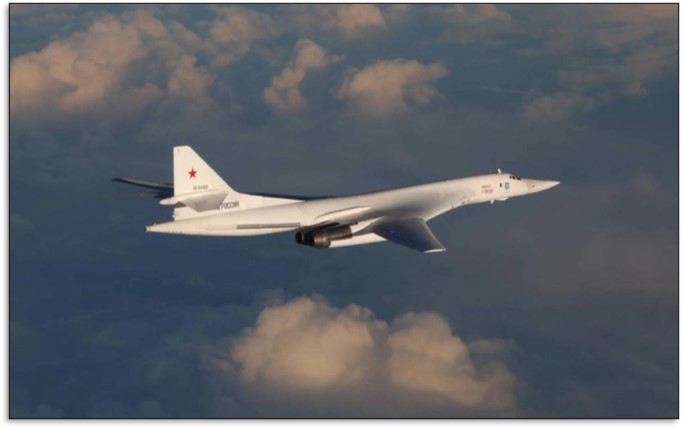}
\caption{A Tu-160 Blackjack strategic bomber identified by Norwegian F-16s on QRA alert in October 2021. PHOTO: Forsvaret and \faktiskbar.}
\label{fig:tu160}
\end{figure*}

The deployment of Russian bombers on the Kola Peninsula in North-Western part of Russia garnered international attention on September 30, 2022, following a report from \textbf{The Jerusalem Post}. The Israeli newspaper's satellite images, from the months of August and September, revealed the presence of four Tu-160 aircraft and three Tu-95s. 
Through the utilisation of satellite imagery obtained from Planet Labs, \faktiskbar conducted a comprehensive analysis of the Olen'ya airfield situated in the North-Western region of Russia's Kola Peninsula. The examination of these satellite images provided valuable insights into the presence of strategic bombers in the area, indicating a long-term and substantial deployment. Notably, these aircraft are typically stationed at the Engels airbase, which is located approximately 720 kilometers southeast of Moscow. To gain a visual understanding, please refer to Figure~\ref{fig:bombersatborders}.

On October 07, 2022, \faktiskbar utilised Planet Labs satellite imagery to capture a significant development at an airfield located approximately 200 kilometers from the Norwegian border. The detailed satellite image clearly displayed the presence of Russian air weaponry at the site. Specifically, the image revealed the presence of seven Tu-160 Blackjack aircraft (shown in Figure~\ref{fig:tu160}) and four Tu-95 Bear-H aircraft parked on the airfield. Subsequent photographs taken two days later showcased one of the Tu-160 bombers positioned for takeoff on the runway. The satellite images unveiled an alarming concentration of long-range strategic bombers stationed at a Russian airbase in close proximity to the Norwegian border. These aircraft possess the capability to launch nuclear bomb attacks against targets in both the United States and Europe.


After conducting further examination of historical satellite imagery obtained from Planet Labs, \faktiskbar uncovered significant findings. It was revealed that the initial group of four Tu-160 aircraft had been stationed at the location since as early as August 21, 2022. Subsequently, on September 25, 2022, three Tu-95 aircraft were also observed at the same location. Numerous satellite images acquired during this period depicted the planes either on the runway or taxiway, indicating preparations for take-off.

According to Ukrainian military intelligence, it has been reported that several strategic bombers deployed at the Kola Peninsula are being used to target Ukrainian locations. On October 8, it was observed that seven Tu-160 bombers at the Olen'ya airfield were armed with Kh-101 cruise missiles, as stated on the agency's official website~\footnote{https://tinyurl.com/mrup5mhc}.

On October 12, 2022, \faktiskbar published this story, which swiftly gained widespread attention and became the headline news in leading online newspapers in Nordic countries. It also captured the interest of international media outlets, including British tabloids, and was featured as a fact-check in Newsweek. The story even made its way to Ukraine, where it received extensive coverage.

\begin{figure*}[t!]
    \centering
    \includegraphics[width=0.9\linewidth]{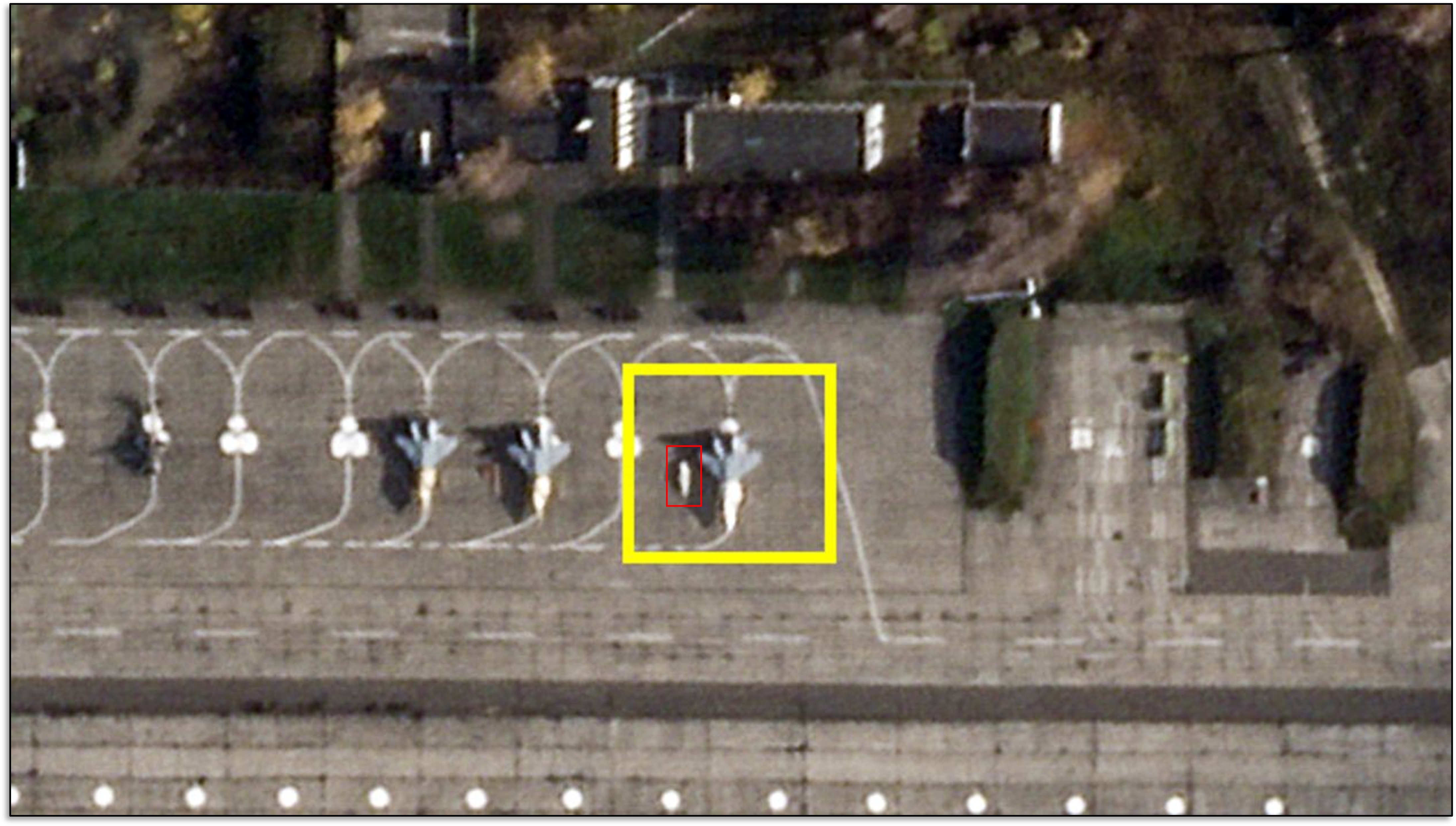}
    \caption{At one of the MiG-31K aircraft (marked with yellow box) that was photographed by a Planet Labs satellite on Sunday morning, there is a white object that is obviously a Kinzjal missile (marked with red box). PHOTO: Planet Labs and \faktiskbar.}
\label{fig:migsinminsk}
    \includegraphics[width=0.6\linewidth]{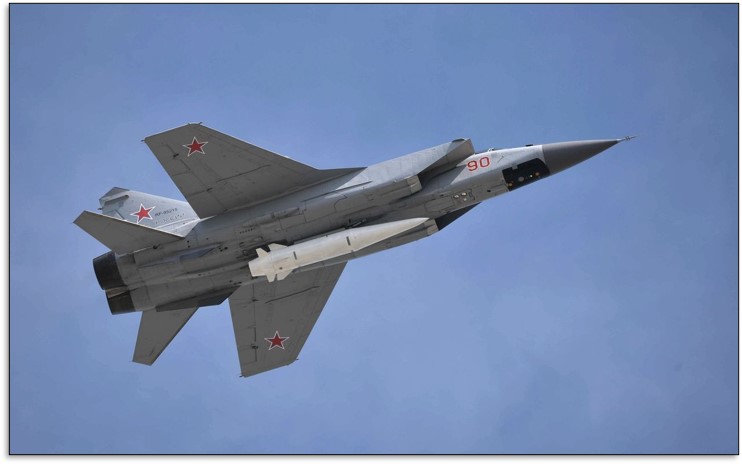}
    \caption{A MiG-31K aircraft with the AS-24 KILLJOY missile attached to its belly. }
\label{fig:migwithkinzal}
\end{figure*}

\subsection{Russian military aircraft deployed in Belarus}

On November 01, 2022, the British Ministry of Defence announced that Russia had deployed its military assets at Machulishchi Airfield in Belarus, near the city of Minsk. The deployed assets reportedly included Russian \textit{MiG-31K FOXHOUND} fighter jets and the highly advanced hyper-sonic Kinzjal missiles, also known as \textit{AS-24 KILLJOY}\footnote{NATO reporting name. List available at: https://tinyurl.com/3fz92c8n}. These missiles have the capability to travel at speeds 10 times the speed of sound and have a range of 2000 kilometers.

To further investigate the story, \faktiskbar obtained recent satellite imagery from Planet Labs, as depicted in Figure~\ref{fig:migsinminsk}. The acquired images revealed the presence of three Russian \textit{MiG-31K FOXHOUND} aircraft on the ground at Machulishchi Airfield, located south of Minsk, on Monday, October 31, 2022, at 09:41 AM. The satellite imagery was captured by Planet Labs' satellites on the previous Sunday morning, a day before \faktiskbar initiated its investigations. Notably, one of the aircraft in the image displayed a clearly visible \textit{Kinzjal} missile positioned on the ground next to it. For visual reference, a photograph featuring a MiG-31K carrying a \textit{Kinzjal} missile is presented in Figure~\ref{fig:migwithkinzal}.

In order to further track the presence of Russian aircraft in Belarus, \faktiskbar obtained historical satellite imagery of the Machulishchi Airfield from Planet Labs. Analysis of the acquired imagery confirmed that the aircraft had been present at the airfield as early as October 18, 2022.

\begin{figure*}[t!]
  \centering
\includegraphics[width=1\linewidth]{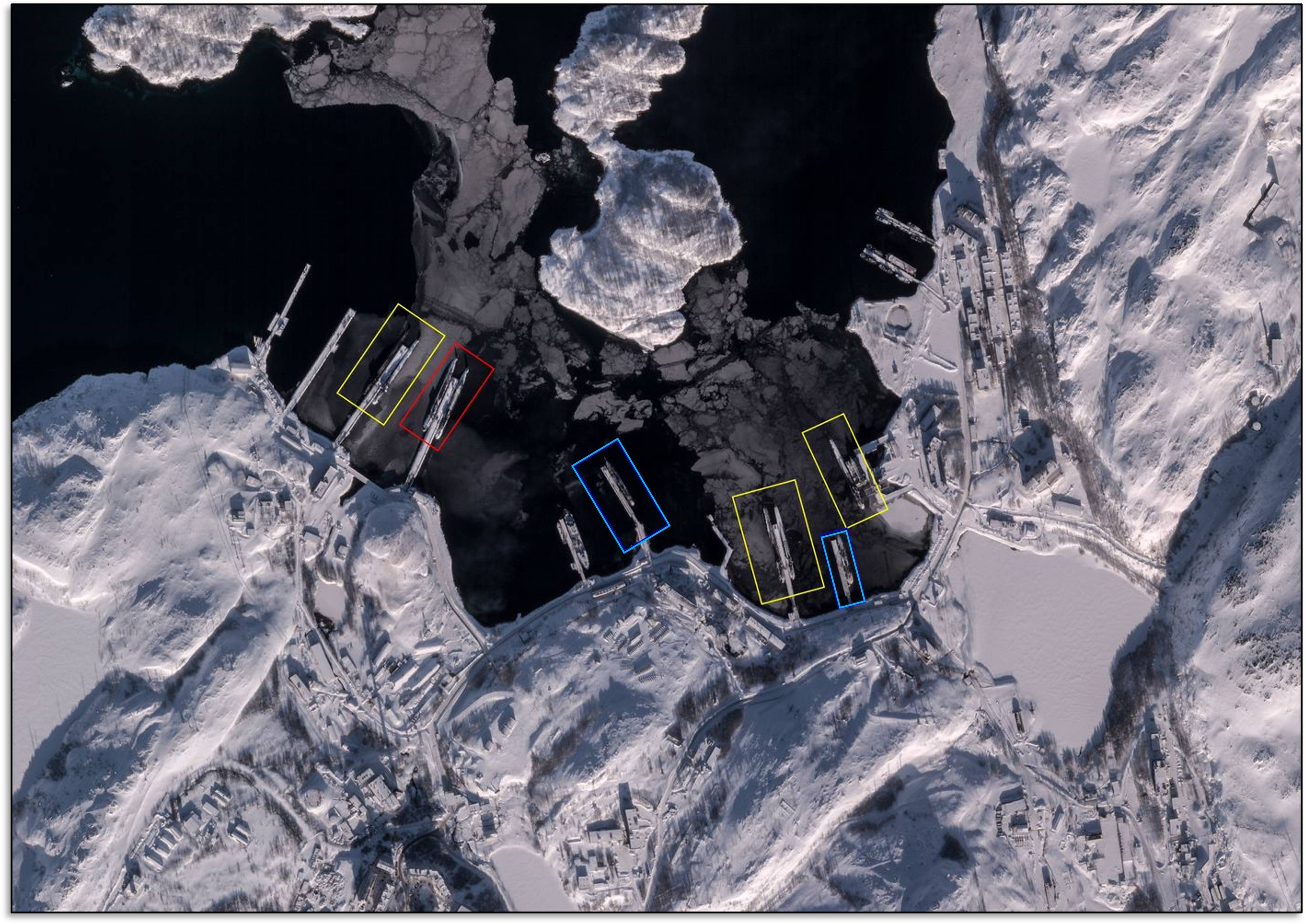}
\caption{A satellite image taken on March 7 of the Gadzhijevo submarine base on the Kola Peninsula. The Akula submarines are marked with blue squares, Delta IV with yellow and Borej with red. PHOTOS: Planet Labs and \faktiskbar.}
\label{fig:submarines}
\end{figure*}

\subsection{Russian nuclear submarines near Norway}
In March 2022, \faktiskbar obtained and examined satellite images of a significant submarine base located on the western side of the Murmansk Fjord, merely 100 kilometers from the Norwegian border. Please refer to Figure~\ref{fig:submarines} for visual representation.

On March 7, 2022, \faktiskbar obtained a satellite image which revealed significant findings at the submarine base. The image depicted the presence of two Borej-class submarines, five Delta IV-class submarines, and two smaller Akula submarines. It is worth noting that Russia heavily relies on its Borej submarines for nuclear deterrence, as they possess the capability to launch strategic missiles that can reach targets on the opposite side of the northern hemisphere within a mere 20 minutes. Among the submarines in the Northern Fleet, two Borej submarines stand out. The Jurij Dolgorukij (K-535) was the first fourth-generation Borej-class submarine to become operational, while the Knyaz Vladimir (K-549) represents an enhanced version (Borej-A) known for its quieter and more stealthy characteristics when compared to other Borej submarines.

As stated by an expert\footnote{Thomas Nilsen: editor at The Independent Barents Observer.}, it is customary for at least two ballistic missile submarines to be operational in the eastern Barents Sea or beneath the ice in the northern region of Novaya Semlja. However, Nilsen suggests that the presence of seven ballistic missile submarines simultaneously docked in the same location, as observed in the image, signifies Russia's intention to maintain a reduced level of tension regarding its 
strategic nuclear forces in the weeks following the invasion of Ukraine.

\section{Lessons Learned and Research Vision}
\label{sec:vision}

\subsection{Tools used at \faktiskbar}
\label{sec:faktisktools}

In the preceding sections, we have detailed \faktiskbar's comprehensive approach to multimedia verification within the context of the Russia-Ukraine conflict. This approach involves the combination of both conventional and specialised OSINT (Open-Source Intelligence) tools. These tools are instrumental in fulfilling various verification needs, encompassing functionalities such as reverse image searches, facial recognition, geolocation, and person identification. 

We have summarised several methods for deducing the date and time when the multimedia content was created or captured in Table~\ref{tab:when}. Additionally, Table~\ref{tab:tools} offers a concise overview of the primary tools consistently employed by \faktiskbar in the pursuit of multimedia content verification. These tools represent a pivotal part of the verification workflow, contributing significantly to the accuracy and reliability of our investigative processes. 

\begin{table*}[t!]
  \centering
  \caption{Useful methods of finding out time and date of the events being presented in images/videos. In the left column we present the methods, whereas on the right side we write a brief description about each method.}
  \resizebox{1.0\linewidth}{!}{%
    \begin{tabular}{P{2.5cm}p{14.5cm}}
    \toprule
    \textbf{Method} & \textbf{Description} \\
    \midrule
    Metadata &
      Nearly all the digital photos/videos have embedded metadata information including the date and time the photo/video was captured among other details. This metadata information can be examined using most freely available photo editing software or online tools.
      \\
    \midrule
    Check the upload date &
      Most social media platforms show the date and time a photo/video was shared along with other details such as, name of the person who shared the post, and on some platforms the location from where the post was shared. This information can also sometimes lead to the date and time at which the photo/video was actually captured.
      \\
    \midrule
    Check for context clues &
      Look for other visible elements in the photo/video that can be helpful in verifying the time and date, such as the sun’s position, weather details (rain, snow), presence of people, or clothing people are wearing.
      \\
    \midrule
    Look for visible time stamps &
      Digital devices such as some video cameras, security cameras etc embed a visible time stamp in the footage which can also be helpful.
      \\
    \midrule
    Compare with other sources &
      If the photo/video was taken/shared in the course of a specific event (festival, bombing), compare it with other photos/videos taken/shared at the same event. This can also sometimes lead to the exact time and date of the acquisition of a photo/video.
      \\
    \midrule
    Analyse weather data &
      Historical weather data can help in estimating the date/time of an event being presented in an image/video. As mentioned in the case above, \faktiskbar was able to roughly estimate the date of a highly important event by analysing the historical weather patterns using Weather.com. For reference, please see Figure~\ref{fig:iskander_weather}.
      \\
    \midrule
    Analyse shadows &
       When there is no metadata information available, or any other contextual clues about the photo/video, the length of shadows can be utilised to establish the time of day when a photo/video was captured. Shadows are longest during the early morning hours and late afternoon, and shortest at noon. To determine the time, measure the length of the shadow cast by a known object inside the photo/video, and then compare it to the object's height. The ratio of the shadow’s length to the object’s height can then be utilised to estimate the time of the day. SunCalc is a freely available online tool which helps in inferring the time at which a photo/video was captured using shadow analysis.
      \\
    \midrule
    Contact the user &
      If still unsure of the time and date, one can try to reach out to the user who shared the photo/video online, and ask for more information.
      \\
      \bottomrule
    \end{tabular}%
    }
  \label{tab:when}%
\end{table*}%

\begin{table*}[t!]
  \centering
  \caption{Frequently used computational tools utilised by \faktiskbar for verifying visuals relating to the conflict in Ukraine.}
    \begin{tabular}{P{3cm}P{10cm}}
    \toprule
    \textbf{Application} & \textbf{Tool}
      \\
    \midrule
    Reverse image search &
      Google Images, Google lens, Yandex Images, Bing Images, Tineye
      \\
    \midrule
    Facial recognition &
      PimEyes(Paid service), Search4faces.com
      \\
    \midrule
    Image verification &
      ImgOps, InVid-WeVerify, FotoForensics
      \\
    \midrule
    Social media &
      Facebook, Twitter, Telegram, TikTok, VKontakte
      \\
    \midrule
    Social media monitoring &
      Dataminr, Tweetdeck
      \\
    \midrule
    Geolocation &
      Google Maps/Streetview, Yandex Maps/Streetview, Bing Maps, Apple Maps
      \\
    \midrule
    Chronolocation &
      NASA FIRMS, SunCalc, Yr.no, Weather.com
      \\
    \midrule
    Satellite imagery &
      NASA FIRMS, Google Earth Pro, Sentinel Hub Playground, Planet.com
      \\
    \midrule
    Workflow &
      Slack and Google sheets, docs, drive, and Gmail
      \\
    \midrule
    Publishing platform &
      NTB Mediebank
      \\
    \bottomrule
    \end{tabular}%
  \label{tab:tools}%
\end{table*}%

\subsection{Lessons Learned and Challenges Ahead}
Multimedia Verification is a distinct field, though it shares similarities with multimedia forensics, investigative journalism, and fact-checking. However, it has unique characteristics, primarily driven by time constraints and limited resources. Here, we will explore the lessons we have learned and the challenges we foresee in the future.

\begin{itemize}[noitemsep, leftmargin=*, wide=0pt]
    \item \textbf{Time Constraints and Resource Limitations:} Multimedia Verification faces tight time limits, often needing to verify the multimedia content in just one day (as shown in 59 out of 81 cases in our study, as described in Figure~\ref{fig:persons_time}). 
    The limited time makes it harder because we do not have a lot of resources. Unlike investigative work or fact-checking, \textit{multimedia verification almost always suffers from a lack of sufficient data}. For instance, when checking a new event, it is common for reverse image search engines not to find similar images. Typically, these engines can retrieve the correct images just two days after the event, but unfortunately, that is too late for verification. This also poses various challenges for machine learning-based verification systems, both in training and validation. 
    
    \item \textbf{Complex Verification Results:} Multimedia verification results are not just black and white, as some might assume. Instead of simply labelling content as true or false, verification outcomes are typically documented in comprehensive reports. These reports offer a thorough evaluation of the content's status, often utilising categories like (a) Verified, (b) Unverifiable, (c) Partially verified, (d) False, or (e) Misleading (refer to Figure~\ref{fig:categories} and Section~\ref{sec:report_fields}). This approach underscores the significance of producing high-quality reports swiftly, as it plays a pivotal role in ensuring the effectiveness of the verification process. However, achieving this balance between quality and speed becomes increasingly challenging due to resource constraints and time limitations.

    \item \textbf{Importance of Human Judgement:} Human element plays a crucial role in multimedia verification. Fact-checkers and verification professionals provide the essential human judgement needed to assess context, evaluate nuances, and make final determinations. In almost all of the cases discussed in this study highlight the importance of the human judgement. This human-in-the-loop approach complements automated processes and ensures a holistic verification process.

   \item \textbf{Multidisciplinary Complexity:} Multimedia verification is inherently complex, requiring expertise from various domains. Verification professionals must draw from fields such as linguistics, image analysis, data mining, journalism, and many others. It is a true collaboration of multidisciplinary knowledge, and success in this field relies on the ability to integrate insights and techniques from these diverse areas. 

Furthermore, navigating collaborations and identifying mutual benefits can present substantial challenges. For instance, while the research community endeavours to produce new knowledge and deepen its understanding of a subject, the media industry is primarily focused on crafting engaging content that generates revenue\cite{opdahl2023trustworthy}. This disparity underscores the need for effective bridging between these two worlds within the context of multimedia verification.

   \item \textbf{Lack of User-Friendly Tools and Tool Limitations:} Multimedia verification grapples with two major issues concerning tools. Firstly, there's a shortage of user-friendly tools, which hinders progress in the field, despite notable advancements in machine learning and computer vision research. For example, \faktiskbar's use of VLC to extract video frames and manually matching interest points (as illustrated in Figure~\ref{fig:bombersatborders}) highlights instances where tasks, long addressed in the domain of computer vision for nearly two decades, could be facilitated with more accessible and modern tools. 
    Secondly, when tools are available, they often come with limitations. These limitations encompass accuracy concerns, difficulties in handling various types of media, and the potential for both false positives and false negatives. For example, let's consider some specific tools:
    \begin{itemize}[noitemsep, leftmargin=*]
    \item \textbf{PimEyes} is a facial search engine that can identify individuals based on their facial characteristics. However, unlike Clearview, which is exclusively accessible to law enforcement, PimEyes has limitations. It cannot search for faces specifically on social media platforms, and its accuracy can vary based on the complexity of the faces it analyses.
    \item \textbf{Dataminr} is a real-time event monitoring via news and social media sources, exhibits different levels of efficiency depending on the language, country, or region of application. Interestingly, it appears to be less efficient when applied to Norwegian news, for reasons that are yet to be determined.
    \item \textbf{NASA FIRMS.} While essential for reporting at \faktiskbar, but NASA FIRMs is constrained by satellite movements. In this study, it was a reliable news service for tracking fires in Ukraine, while other times it has failed to detect fires that have been confirmed by other sources. The effectiveness of FIRMS is largely dependent on the satellite's proximity to the target area and the duration of the fire. 
    \end{itemize}

    \item \textbf{Big Tech Influence:} The influence of major technology companies, e.g., Google, Meta or Microsoft, in multimedia verification is undeniable. These companies control over crucial resources, algorithms, and platforms, such as reverse image search engines. This influence can create disparities in access and capabilities among different verification entities. Achieving a more equitable distribution of resources and reducing reliance on tech giants is imperative for the field's sustainability.

    \item \textbf{Limited Research and Generative AI Challenges:} 
    Despite making good progress, the field of multimedia verification still lacks thorough research and consistent evaluation standards. While initiatives like the Multimedia Verification tasks at MediaEval~\cite{boididou2015verifying, boididou2018verifying} or the grand challenge on detecting cheapfakes~\cite{dang2023grand}, have expanded the field, there remains a need for more extensive and ongoing research.
    
    Furthermore, the rise of generative AI, for example deepfakes~\cite{Tolosana2020DeepFakesAB}, has introduced a new dimension to the challenge of combating mis/disinformation. And now, the introduction of Diffusion Models~\cite{Yang2022DiffusionMA} has made the generation of convincing synthetic content considerably easy. These new AI models have raised the bar for verifying the legitimacy and authenticity of multimedia content~\cite{techwireasia}. This presents a significant obstacle for fact-checkers and journalists who lack dedicated tools capable of effectively detecting and verifying AI-generated synthetic content. As a result, there is a pressing need for the development of advanced computational tools that can accurately identify and authenticate such content, enabling fact-checkers and journalists to address the growing threat of generative AI misinformation and ensure the dissemination of accurate information to the public. 
\end{itemize}

Additional considerations: Beyond these core lessons, it is crucial to address factors such as data quality and availability, real-time challenges, ethical considerations, cross-platform verification, user training and education, international and cross-linguistic challenges, collaboration, and the development of open-source resources. These considerations ensure a comprehensive and adaptable approach to multimedia verification.

\subsection{Research Vision for Multimedia Verification}
Drawing from the insights gained and our anticipation of forthcoming challenges, we present the following directions in multimedia verification:

\begin{itemize}[noitemsep, leftmargin=*, wide=0pt]
\item \textbf{Real-Time Verification:} Researchers should focus on developing advanced real-time verification solutions capable of handling multimedia content within a very short time-frame, even in the face of limited resources. This would involve harnessing the power of machine learning, edge computing, and distributed systems to meet the stringent time constraints of the digital information ecosystem.

\item \textbf{Enhanced Multimodal Analysis:} Future research should emphasise the integration of multiple types of content analysis, including text, images, videos, and audio. Advancements in natural language processing, computer vision, multimedia forensics, and audio processing should converge to enable more comprehensive and accurate verification results. 

\item \textbf{AI-Human Collaboration Models:} Explore innovative AI-human collaboration models that leverage the strengths of both machines and human experts. Develop AI systems that can seamlessly integrate human judgement and domain expertise, creating synergy in the verification process. For example, deep learning models can be utilised to empower journalists in efficiently and automatically analysing/detecting synthetic content generated by deep neural network models~\cite{Tolosana2020DeepFakesAB}. These models can be added to tools like InVid-WeVerify verification plugin~\cite{MezarisInVid}, making it easier for people who verify the multimedia content.

\item \textbf{Explainable AI for Verification:} Develop explainable AI models that can provide clear justifications for verification decisions. This is crucial for building trust in automated verification systems, as it allows users to understand why a piece of content is classified as true, false, misleading, or simply unverifiable.

\item \textbf{Data Augmentation Strategies:} Given the lack of data for training, researchers should explore data augmentation techniques to expand the effectiveness of machine learning models. Synthetic data generation and transfer learning can help mitigate data limitations. 

\item \textbf{Open Source Verification Tools:} Create and maintain open-source verification tools, for example FotoVerifier~\cite{tran2022dedigi}, that are user-friendly and accessible. Encourage collaboration among researchers, journalists, and fact-checkers to continuously improve and refine these tools for widespread adoption.

\item \textbf{Ethical AI and Bias Mitigation:} Prioritise research on ethical AI in multimedia verification. Develop methods to identify and mitigate biases in verification algorithms to ensure fair and unbiased content assessment.

\item \textbf{Interdisciplinary Training Programs:} Establish interdisciplinary training programs that equip verification professionals with expertise in machine learning, media forensics, linguistics, image analysis, data mining, and journalism. This holistic approach will better prepare individuals for the complex nature of multimedia verification.

\item \textbf{Standardised Evaluation Benchmarks:} Collaborate on the creation of standardised evaluation benchmarks and datasets that reflect real-world multimedia verification challenges, for examples, the Multimedia Verification tasks at MediaEval~\cite{boididou2015verifying} and the grand challenge in detecting cheapfakes~\cite{dang2023grand}. These benchmarks should be continuously updated to keep pace with evolving threats and media formats. 

\item \textbf{Collaboration with Tech Giants:} Engage in partnerships and dialogues with major technology companies to promote transparency and resource-sharing in the field. Encourage these companies to provide access to their tools and algorithms to ensure equitable distribution of resources. 

\item \textbf{Counter Generative AI Threats:} Invest in research to counter the evolving threats posed by generative AI. Develop advanced detection techniques that can identify increasingly convincing fake multimedia content generated by deep learning models.

\item \textbf{User Education and Media Literacy:} Promote research on user education and media literacy programs (for example, the Digital Information Literacy Guide~\cite{kari2022literacy}). Educating the public about the challenges of multimedia verification and critical thinking skills is crucial for combating misinformation at its source. 
\end{itemize}

\section{Conclusion}

In conclusion, this study showcases the vital role of multimedia verification in combating misinformation during pivotal events. It highlights how computational tools and Open-Source Intelligence (OSINT) techniques empower journalists and fact-checkers to authenticate information from various online sources, ensuring the dissemination of reliable and accurate information to the public.

Specifically focusing on \faktiskbar's efforts during the Russia-Ukraine conflict, this paper explores their unique workflows and the use of OSINT techniques and computational tools for verifying news stories from the conflict zone. We have detailed case studies spanning nine months, from April to December 2022, providing a comprehensive understanding of \faktiskbar's verification methodologies.

As we conclude this research, we anticipate further exploration and innovation in the field of multimedia verification, seeking to continually enhance our ability to combat misinformation and promote the dissemination of trustworthy information in our ever-evolving digital landscape.

\section*{Acknowledegments}
This research was supported by the NORDIS, European Horizon 2020 grant number 825469; and by industry partners and the Research Council of Norway with funding to MediaFutures: Research Centre for Responsible Media Technology and Innovation, through the Centres for Research-based Innovation scheme, project number 309339.

\section*{Declaration of generative AI and AI-assisted technologies in the writing process
}
During the preparation of this work the author(s) used ChatGPT~\cite{chatgpt} in order to check and correct the grammar of this paper. After using this tool/service, the author(s) reviewed and edited the content as needed and take(s) full responsibility for the content of the publication.



\printcredits

\bibliographystyle{cas-model2-names}

\bibliography{ref}

\begin{thebibliography}{24}
\expandafter\ifx\csname natexlab\endcsname\relax\def\natexlab#1{#1}\fi
\providecommand{\url}[1]{\texttt{#1}}
\providecommand{\href}[2]{#2}
\providecommand{\path}[1]{#1}
\providecommand{\DOIprefix}{doi:}
\providecommand{\ArXivprefix}{arXiv:}
\providecommand{\URLprefix}{URL: }
\providecommand{\Pubmedprefix}{pmid:}
\providecommand{\doi}[1]{\href{http://dx.doi.org/#1}{\path{#1}}}
\providecommand{\Pubmed}[1]{\href{pmid:#1}{\path{#1}}}
\providecommand{\bibinfo}[2]{#2}
\ifx\xfnm\relax \def\xfnm[#1]{\unskip,\space#1}\fi
\bibitem[{Boididou et~al.(2015)Boididou, Andreadou, Papadopoulos, Dang~Nguyen,
  Boato, Riegler, Kompatsiaris et~al.}]{boididou2015verifying}
\bibinfo{author}{Boididou, C.}, \bibinfo{author}{Andreadou, K.},
  \bibinfo{author}{Papadopoulos, S.}, \bibinfo{author}{Dang~Nguyen, D.T.},
  \bibinfo{author}{Boato, G.}, \bibinfo{author}{Riegler, M.},
  \bibinfo{author}{Kompatsiaris, Y.}, et~al., \bibinfo{year}{2015}.
\newblock \bibinfo{title}{{Verifying Multimedia Use at Mediaeval 2015}}, in:
  \bibinfo{booktitle}{{MediaEval 2015}}. \bibinfo{publisher}{CEUR-WS}. volume
  \bibinfo{volume}{1436}.
\bibitem[{Boididou et~al.(2018)Boididou, Middleton, Jin, Papadopoulos,
  Dang-Nguyen, Boato and Kompatsiaris}]{boididou2018verifying}
\bibinfo{author}{Boididou, C.}, \bibinfo{author}{Middleton, S.E.},
  \bibinfo{author}{Jin, Z.}, \bibinfo{author}{Papadopoulos, S.},
  \bibinfo{author}{Dang-Nguyen, D.T.}, \bibinfo{author}{Boato, G.},
  \bibinfo{author}{Kompatsiaris, Y.}, \bibinfo{year}{2018}.
\newblock \bibinfo{title}{{Verifying Information with Multimedia Content on
  Twitter: A Comparative Study of Automated Approaches}}.
\newblock \bibinfo{journal}{{Multimedia tools and applications}}
  \bibinfo{volume}{77}, \bibinfo{pages}{15545--15571}.
\bibitem[{Cherkasets(2019)}]{osint_team_tools}
\bibinfo{author}{Cherkasets, P.}, \bibinfo{year}{2019}.
\newblock \bibinfo{title}{{OSINT: How to Find Information on Anyone}}.
\newblock \bibinfo{howpublished}{URL: https://tinyurl.com/vx94c8j4}.
\bibitem[{Corcoran(2018)}]{reuters-satellite}
\bibinfo{author}{Corcoran, M.}, \bibinfo{year}{2018}.
\newblock \bibinfo{title}{{Satellite Journalism – The Big Picture}}.
\newblock \bibinfo{howpublished}{\url{https://tinyurl.com/5n7ucchf}}.
\bibitem[{Dang-Nguyen et~al.(2023)Dang-Nguyen, Khan, Midoglu, Riegler,
  Halvorsen and Dao}]{dang2023grand}
\bibinfo{author}{Dang-Nguyen, D.T.}, \bibinfo{author}{Khan, S.A.},
  \bibinfo{author}{Midoglu, C.}, \bibinfo{author}{Riegler, M.},
  \bibinfo{author}{Halvorsen, P.}, \bibinfo{author}{Dao, M.S.},
  \bibinfo{year}{2023}.
\newblock \bibinfo{title}{{Grand Challenge On Detecting Cheapfakes}}.
\newblock \bibinfo{journal}{arXiv e-prints} , \bibinfo{pages}{arXiv--2304}.
\bibitem[{Gill(2023)}]{osint_definition}
\bibinfo{author}{Gill, R.}, \bibinfo{year}{2023}.
\newblock \bibinfo{title}{{What is Open-Source Intelligence?}}
\newblock \bibinfo{howpublished}{URL: https://tinyurl.com/3hwf6rdm}.
\bibitem[{ICFJ(2019)}]{icfjreport2019}
\bibinfo{author}{ICFJ}, \bibinfo{year}{2019}.
\newblock \bibinfo{title}{{State of Technology in Global Newsrooms:
  International Centre For Journalists (ICFJ)}}.
\newblock \bibinfo{howpublished}{URL: https://tinyurl.com/yc7cv2j5}.
\bibitem[{Ireton and Posetti(2018)}]{Ireton2018JournalismFN}
\bibinfo{author}{Ireton, C.}, \bibinfo{author}{Posetti, J.},
  \bibinfo{year}{2018}.
\newblock \bibinfo{title}{{Journalism, Fake News \& Disinformation}}.
\bibitem[{Khan et~al.(2023)Khan, Sheikhi, Opdahl, Rabbi, Stoppel, Trattner and
  Dang-Nguyen}]{khan2023visual}
\bibinfo{author}{Khan, S.A.}, \bibinfo{author}{Sheikhi, G.},
  \bibinfo{author}{Opdahl, A.}, \bibinfo{author}{Rabbi, F.},
  \bibinfo{author}{Stoppel, S.}, \bibinfo{author}{Trattner, C.},
  \bibinfo{author}{Dang-Nguyen, D.T.}, \bibinfo{year}{2023}.
\newblock \bibinfo{title}{{Visual User-Generated Content Verification in
  Journalism: An Overview}}.
\newblock \bibinfo{journal}{IEEE Access} \bibinfo{volume}{11},
  \bibinfo{pages}{6748--6769}.
\bibitem[{Kivinen et~al.(2022)Kivinen, Horowitz, Havula, Härkönen, Kiili,
  Kivinen, Pönkä, Pörsti, Salo, Vuorikari and Vahti}]{kari2022literacy}
\bibinfo{author}{Kivinen, K.}, \bibinfo{author}{Horowitz, M.A.},
  \bibinfo{author}{Havula, P.}, \bibinfo{author}{Härkönen, T.},
  \bibinfo{author}{Kiili, C.}, \bibinfo{author}{Kivinen, E.},
  \bibinfo{author}{Pönkä, H.}, \bibinfo{author}{Pörsti, J.},
  \bibinfo{author}{Salo, M.}, \bibinfo{author}{Vuorikari, R.},
  \bibinfo{author}{Vahti, J.}, \bibinfo{year}{2022}.
\newblock \bibinfo{title}{{Digital Information Literacy Guide: A Digital
  Information Literacy Guide for Citizens in the Digital Age}} .
\bibitem[{Mossou and Higgins(2021)}]{bellingcat2021verificationguide}
\bibinfo{author}{Mossou, A.}, \bibinfo{author}{Higgins, R.},
  \bibinfo{year}{2021}.
\newblock \bibinfo{title}{{A Beginner's Guide to Social Media Verification}}.
\newblock \bibinfo{howpublished}{URL: https://tinyurl.com/mr4y5edv}.
\bibitem[{Opdahl et~al.(2023)Opdahl, Tessem, Dang-Nguyen, Motta, Setty,
  Throndsen, Tverberg and Trattner}]{opdahl2023trustworthy}
\bibinfo{author}{Opdahl, A.L.}, \bibinfo{author}{Tessem, B.},
  \bibinfo{author}{Dang-Nguyen, D.T.}, \bibinfo{author}{Motta, E.},
  \bibinfo{author}{Setty, V.}, \bibinfo{author}{Throndsen, E.},
  \bibinfo{author}{Tverberg, A.}, \bibinfo{author}{Trattner, C.},
  \bibinfo{year}{2023}.
\newblock \bibinfo{title}{{Trustworthy Journalism through AI}}.
\newblock \bibinfo{journal}{Data \& Knowledge Engineering}
  \bibinfo{volume}{146}, \bibinfo{pages}{102182}.
\bibitem[{OpenAI(2021)}]{chatgpt}
\bibinfo{author}{OpenAI}, \bibinfo{year}{2021}.
\newblock \bibinfo{title}{{Introducing ChatGPT}}.
\newblock \bibinfo{howpublished}{\url{https://openai.com/blog/chatgpt}}.
\bibitem[{Pasquini et~al.(2021)Pasquini, Amerini and Boato}]{Pasquini2021}
\bibinfo{author}{Pasquini, C.}, \bibinfo{author}{Amerini, I.},
  \bibinfo{author}{Boato, G.}, \bibinfo{year}{2021}.
\newblock \bibinfo{title}{{Media Forensics on Social Media Platforms: A
  Survey}}.
\newblock \bibinfo{journal}{EURASIP Journal on Information Security}
  \bibinfo{volume}{2021}, \bibinfo{pages}{1 -- 19}.
\bibitem[{Raj(2023)}]{techwireasia}
\bibinfo{author}{Raj, A.}, \bibinfo{year}{2023}.
\newblock \bibinfo{title}{{Deepfakes Detection Gets Tougher}}.
\newblock \bibinfo{howpublished}{\url{https://tinyurl.com/2p8nrssc}}.
\bibitem[{Schifferes et~al.(2014)Schifferes, Newman, Thurman, Corney, G{\"o}ker
  and Martin}]{Schifferesnewsverification}
\bibinfo{author}{Schifferes, S.}, \bibinfo{author}{Newman, N.},
  \bibinfo{author}{Thurman, N.J.}, \bibinfo{author}{Corney, D.},
  \bibinfo{author}{G{\"o}ker, A.}, \bibinfo{author}{Martin, C.},
  \bibinfo{year}{2014}.
\newblock \bibinfo{title}{{Identifying and Verifying News Through Social
  Media}}.
\newblock \bibinfo{journal}{Digital Journalism} \bibinfo{volume}{2},
  \bibinfo{pages}{406 -- 418}.
\bibitem[{Silverman(2013)}]{Silverman2013VerificationH}
\bibinfo{author}{Silverman, C.L.}, \bibinfo{year}{2013}.
\newblock \bibinfo{title}{{Verification Handbook : An Ultimate Guideline on
  Digital Age Sourcing for Emergency Coverage}}.
\bibitem[{Teyssou et~al.(2017)Teyssou, Leung, Apostolidis, Apostolidis,
  Papadopoulos, Zampoglou, Papadopoulou and Mezaris}]{MezarisInVid}
\bibinfo{author}{Teyssou, D.}, \bibinfo{author}{Leung, J.M.},
  \bibinfo{author}{Apostolidis, E.}, \bibinfo{author}{Apostolidis, K.},
  \bibinfo{author}{Papadopoulos, S.}, \bibinfo{author}{Zampoglou, M.},
  \bibinfo{author}{Papadopoulou, O.}, \bibinfo{author}{Mezaris, V.},
  \bibinfo{year}{2017}.
\newblock \bibinfo{title}{{The InVID Plug-in: Web Video Verification on the
  Browser}}.
\newblock \bibinfo{journal}{Proceedings of the First International Workshop on
  Multimedia Verification} .
\bibitem[{Thomson et~al.(2020)Thomson, Angus, Dootson, Hurcombe and
  Smith}]{Thomson2020}
\bibinfo{author}{Thomson, T.J.}, \bibinfo{author}{Angus, D.},
  \bibinfo{author}{Dootson, P.}, \bibinfo{author}{Hurcombe, E.},
  \bibinfo{author}{Smith, A.}, \bibinfo{year}{2020}.
\newblock \bibinfo{title}{{Visual Mis/disinformation in Journalism and Public
  Communications: Current Verification Practices, Challenges, and Future
  Opportunities}}.
\newblock \bibinfo{journal}{Journalism Practice} \bibinfo{volume}{16},
  \bibinfo{pages}{938 -- 962}.
\bibitem[{Tolosana et~al.(2020)Tolosana, Vera-Rodr{\'i}guez, Fierrez, Morales
  and Ortega-Garcia}]{Tolosana2020DeepFakesAB}
\bibinfo{author}{Tolosana, R.}, \bibinfo{author}{Vera-Rodr{\'i}guez, R.},
  \bibinfo{author}{Fierrez, J.}, \bibinfo{author}{Morales, A.},
  \bibinfo{author}{Ortega-Garcia, J.}, \bibinfo{year}{2020}.
\newblock \bibinfo{title}{{DeepFakes and Beyond: A Survey of Face Manipulation
  and Fake Detection}}.
\newblock \bibinfo{journal}{Information Fusion} \bibinfo{volume}{64},
  \bibinfo{pages}{131--148}.
\bibitem[{Tran et~al.(2022)Tran, Tran, Long-Vu, Nguyen, Tran and
  Dang-Nguyen}]{tran2022dedigi}
\bibinfo{author}{Tran, C.H.}, \bibinfo{author}{Tran, Q.T.},
  \bibinfo{author}{Long-Vu, Q.C.}, \bibinfo{author}{Nguyen, H.S.},
  \bibinfo{author}{Tran, A.D.}, \bibinfo{author}{Dang-Nguyen, D.T.},
  \bibinfo{year}{2022}.
\newblock \bibinfo{title}{Dedigi: a privacy-by-design platform for image
  forensics}, in: \bibinfo{booktitle}{Proceedings of the 3rd ACM Workshop on
  Intelligent Cross-Data Analysis and Retrieval}, pp. \bibinfo{pages}{58--62}.
\bibitem[{Turner(2018)}]{Niemanreports}
\bibinfo{author}{Turner, D.}, \bibinfo{year}{2018}.
\newblock \bibinfo{title}{{Truth in the Age of Social Media}}.
\newblock \bibinfo{howpublished}{URL: https://tinyurl.com/2k8j7xbh}.
\bibitem[{Walker and Matsa(2021)}]{pewrresearchreport}
\bibinfo{author}{Walker, M.}, \bibinfo{author}{Matsa, K.E.},
  \bibinfo{year}{2021}.
\newblock \bibinfo{title}{{News Consumption Across Social Media in 2021}}.
\newblock \bibinfo{howpublished}{URL: https://tinyurl.com/yeammb92}.
\bibitem[{Yang et~al.(2022)Yang, Zhang, Hong, Xu, Zhao, Shao, Zhang, Yang and
  Cui}]{Yang2022DiffusionMA}
\bibinfo{author}{Yang, L.}, \bibinfo{author}{Zhang, Z.}, \bibinfo{author}{Hong,
  S.}, \bibinfo{author}{Xu, R.}, \bibinfo{author}{Zhao, Y.},
  \bibinfo{author}{Shao, Y.}, \bibinfo{author}{Zhang, W.},
  \bibinfo{author}{Yang, M.H.}, \bibinfo{author}{Cui, B.},
  \bibinfo{year}{2022}.
\newblock \bibinfo{title}{{Diffusion Models: A Comprehensive Survey of Methods
  and Applications}}.
\newblock \bibinfo{journal}{ArXiv} \bibinfo{volume}{abs/2209.00796}.
\newblock \URLprefix \url{https://api.semanticscholar.org/CorpusID:252070859}.

\end{thebibliography}



\end{document}